\documentclass[preprint,12pt,number,sort&compress]{elsarticle}
\usepackage{amssymb}
\usepackage{xcolor} % For colors
\usepackage{amsmath}
\usepackage{float}
\usepackage[utf8]{inputenc}
\usepackage[T1]{fontenc}
\usepackage{lmodern}
\usepackage{graphicx}
\usepackage[figurename=Fig.,labelfont=bf,labelsep=period]{caption}
\usepackage{subcaption}
\usepackage{amsmath}
\usepackage{amsfonts}
\usepackage{amssymb}
\usepackage{amsbsy}
\usepackage{newtxtext,newtxmath}
\usepackage{subcaption}
\usepackage{tikz}
\usepackage{siunitx} 		    % num()
\usepackage{numprint}		    % thousand separator
\usepackage[normalem]{ulem}		% striking through text horizontally \sout{}
\usetikzlibrary{shapes,arrows}
\usepackage[colorlinks=true,citecolor=blue,linkcolor=magenta]{hyperref}
\usepackage[demo]{adjustbox}
\usepackage{xcolor}
\usepackage{atbegshi}% http://ctan.org/pkg/atbegshi
\AtBeginDocument{\AtBeginShipoutNext{\AtBeginShipoutDiscard}}	% Upper and this line clears first blank page (http://tex.stackexchange.com/questions/140168/how-to-remove-a-blank-page-before-the-title-page)
\PassOptionsToPackage{hyphens}{url}\usepackage{hyperref} % rješava problem dugačkih url linkova.

\begin{document}
	
	\pagestyle{empty} % Removes page numbers
	\begin{titlepage}
		\color[rgb]{.4,.4,1}
		\hspace{5mm}

		\bigskip
		
		\hspace{15mm}
		\begin{minipage}{10mm}
			\color[rgb]{.7,.7,1}
			\rule{1pt}{226mm}
		\end{minipage}
		\begin{minipage}{133mm}
			\vspace{10mm}        
			\color{black}
			\sffamily
			\LARGE\bfseries Physics-informed neural networks   \\[-0.3\baselineskip] for nonlocal beam eigenvalue problems \\[-0.3\baselineskip] 
			
			\vspace{5mm}
			{\large {Preprint of the article published in \\[-0.4\baselineskip] Thin-Walled Structures (2026) }} 
			
			\vspace{10mm}        
			{\large Baidehi Das, Raffaele Barretta, Marko \v{C}ana\dj{}ija } % Author name
			
			\large
			
			\vspace{40mm}
			%		    \small{$^*$ Corresponding author: marko.canadija@riteh.hr. Tel.: +385-51-651-496}	
			\vspace{5mm}
			
			\small
			\url{https://doi.org/10.1016/j.tws.2026.114530}
			
			\textcircled{c} 2026. This manuscript version is made available under the CC-BY-NC-ND 4.0 license \url{http://creativecommons.org/licenses/by-nc-nd/4.0/}
			\hspace{30mm} % or \hfill, if you want the square sticked
			\color[rgb]{.4,.4,1} %                        to the right margin
			\includegraphics[width=3cm]{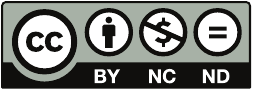}        
		\end{minipage}
	\end{titlepage}
%\begin{frontmatter}

%% Title, authors and addresses

%% use the tnoteref command within \title for footnotes;
%% use the tnotetext command for theassociated footnote;
%% use the fnref command within \author or \affiliation for footnotes;
%% use the fntext command for theassociated footnote;
%% use the corref command within \author for corresponding author footnotes;
%% use the cortext command for theassociated footnote;
%% use the ead command for the email address,
%% and the form \ead[url] for the home page:
%% \title{Title\tnoteref{label1}}
%% \tnotetext[label1]{}
%% \author{Name\corref{cor1}\fnref{label2}}
%% \ead{email address}
%% \ead[url]{home page}
%% \fntext[label2]{}
%% \cortext[cor1]{}
%% \affiliation{organization={},
%%            addressline={}, 
%%            city={},
%%            postcode={}, 
%%            state={},
%%            country={}}
%% \fntext[label3]{}

\title{Physics-Informed Neural Networks for Nonlocal Beam Eigenvalue Problems}

%% use optional labels to link authors explicitly to addresses:
%% \author[label1,label2]{}
%% \affiliation[label1]{organization={},
%%             addressline={},
%%             city={},
%%             postcode={},
%%             state={},
%%             country={}}
%%
%% \affiliation[label2]{organization={},
%%             addressline={},
%%             city={},
%%             postcode={},
%%             state={},
%%             country={}}
% %% Author affiliation
\affiliation[1]{organization={Department of  Civil Engineering and Architecture, University of Catania},
	addressline={Viale Andrea Doria 6}, 
	city={Catania},
	country={Italy}}

\affiliation[2]{organization={Department of Structures for Engineering and Architecture, University of Naples Federico II},%Department and Organization
	addressline={Via Claudio 21}, 
	city={Naples},
	postcode={80125}, 
	country={Italy}}

\affiliation[3]{organization={Faculty of Engineering, University of Rijeka},%Department and Organization
	addressline={Vukovarska 58}, 
	city={Rijeka},
	postcode={51000}, 
	country={Croatia}}
	
\author[1]{Baidehi Das}
\author[2]{Raffaele Barretta}
\author[3]{Marko \v{C}ana\dj{}ija\corref{cor1}} %% Author name

\ead{marko.canadija@riteh.uniri.hr}

\cortext[cor1]{Corresponding author}
%% Abstract
\begin{abstract}
%% Text of abstract
The present study investigates the dynamics of stress-driven nonlocal elastic beams exploiting the Physics-Informed Neural Network (PINN) approach. Specifically, a PINN is developed to compute the first eigenfunction and eigenvalue arising from the underlying sixth-order ordinary differential equation. The PINN is based on a feedforward neural network, with a loss function composed of terms from the differential equation, the normalization condition, and both classical boundary and constitutive boundary conditions. Relevant eigenvalues are treated as separate trainable variables. The results demonstrate that the proposed method is a powerful tool for addressing the complexity of the problem. The obtained results are compared with benchmark analytical solutions and show strong agreement.
\end{abstract}

%% Keywords
%\end{frontmatter}

\maketitle

\section{Introduction}
\label{1}
The design and fabrication of ultrasmall devices such as Micro-Electro-Mechanical Systems (MEMS) and Nano-Electro-Mechanical Systems (NEMS) have a wide range of applications, including energy harvesting \cite{Ren2025}, novel noise reduction devices \cite{LU2024111662}, healthcare monitoring systems \cite{mi16050522}, mechanics of advanced materials \cite{Izadi2024,Yang2024}, and innovative structural designs \cite{Canadija2024,GABRELIAN2024113116}. However, working with these devices at the micro- and nanoscale is challenging, as forces and interactions at the atomistic and molecular levels play a significant role. Although numerous atomistic \cite{Choyal2025,Moradi2025,KUSHCH2023103806} and molecular approaches \cite{jcs8080293,Kumar02072024} have been developed, their solutions are often computationally prohibitive. It is important to note that the computational cost can be drastically reduced by combining molecular dynamics, finite elements, and input convex neural networks (see, e.g., \cite{Canadija2021, Kosmerl2022, Canadija2024a, Canadija2024}) in studies of carbon nanotubes. Another way to reduce the computational burden is through nonlocal continuum theories, which began to appear several decades ago \cite{rogula1965influence, kroner1967elasticity, Rogula1982} and were developed for both static and dynamic problems. These nonlocal elasticity theories have successfully captured size effects that conventional local laws fail to describe.

Eringen \cite{Eringen1972,Eringen19834703} formulated an integral model of elasticity in which the stress field is expressed as a spatial convolution between an appropriate averaging kernel and the elastic strain field. However, as first noted by \cite{peddieson2003application}, the application of Eringen’s approach to structural mechanics led to several inconsistencies and paradoxes. By exploiting the differential form associated with Eringen’s nonlocal model, the absence of scale effects was shown in benchmark cases study \cite{peddieson2003application}. To address these issues, a two-phase nonlocal law was proposed by Eringen himself, id est a convex combination of local and nonlocal approaches. Nevertheless,  the asymptotic response corresponding to the pure nonlocal phase degenerates into an ill-posed formulation. It is worth highlighting that, for a bounded domain, Eringen’s integral law inherently requires the fulfillment of homogeneous constitutive boundary conditions \cite{Polyanin}. This aspect was first discussed in \cite{BENVENUTI201346,Khodabakhshi2015} and finally clarified in \cite{romano2017constitutive}, where it was shown that ill-posedness of the structural problems based on the pure strain-driven Eringen's
nonlocal model stemmed from incompatibility between equilibrium requirements and the constitutive integral law. Notably, it has been shown that the constitutive integral law provided by Eringen admits either a unique solution or no solution at all, depending on whether or not the equilibrated stress fulfills the above mentioned constitutive boundary conditions. With the exception of some special loading cases (see e.g. \cite{Song,Song2}), the constitutive requirements are in contrast with the equilibrium conditions, so that in general the nonlocal elastostatic problem does not admit solution, as shown in \cite{romano2017constitutive} for simple exemplar structural schemes. 
Moreover, it has recently been shown that the obstruction of equilibrium due to Eringen’s integral law emerges not only for bounded domains, but also for unbounded continuum problems, as proven in \cite{unb}.

To overcome this limitation, a consistent stress-driven integral approach was proposed by \cite{romano2017nonlocal,romano2017stress} which has been successfully applied to various static \cite{LOVISI2023117549,altekin2025stress,Barretta01062020} and dynamic \cite{DAS2025119057,Feo16122024,Vaccaro2025} analytical problems. Nevertheless, despite its robustness, solving complex mechanics problems analytically - particularly those involving higher-order differential formulations - remains challenging.

On the other hand, Neural Networks (NNs) have emerged as powerful tools for addressing complex challenges in structural engineering by modeling nonlinear relationships and processing large datasets. By learning from experimental or simulated data, NNs capture complex nonlinear interactions through interconnected nodes, enabling a wide range of engineering applications. Common types of artificial neural networks (ANNs) include Feedforward Neural Networks (FNNs) and Deep Neural Networks (DNNs), where multiple hidden layers enhance feature extraction for complex tasks. Other architectures, such as Convolutional Neural Networks (CNNs) and Deep Belief Networks (DBNs), have also been employed, particularly for probabilistic structural analysis. Similarly, data-driven techniques have caught attention for their ability to learn complex patterns and relationships directly from large datasets, enabling predictions in structural engineering tasks without explicit reliance on governing physical equations, see \cite{samadian2025application, Stocker2024, Zlatic2024, Fuhg2025}.

However, data-driven methods and other approaches discussed above often require large datasets, which can pose a significant computational burden \cite{YUAN2022111260}. A promising way to mitigate this challenge is to integrate established physical knowledge of the system into the learning process. Such prior physical understanding can be embedded in the training procedure by enforcing the residuals of the governing ordinary or partial differential equations (ODEs or PDEs) at selected training points-effectively incorporating the physical laws into the learning process. This approach, known as Physics-Informed Neural Networks (PINNs), was first introduced in \cite{RAISSI2019686}.

There are several advantages of PINNs over classical methods for solving differential equations, such as Runge-Kutta or Finite Element/Finite Difference methods:
\begin{itemize}
	\item In the classical approach, high-dimensional problems are typically solved using basis functions, polynomials, piecewise polynomials, and similar methods. However, these methods are prone to an exponential increase in complexity and computational cost as the number of dimensions grows, a phenomenon commonly referred to as the curse of dimensionality. PINNs, by contrast, have shown greater efficiency in addressing such problems \cite{han2018solving}, although they also face limitations, particularly when handling higher-order derivatives.
	\item Unlike classical numerical methods, PINNs do not suffer from the accumulation of numerical errors over time \cite{Mattheakis2022}, owing to their iterative learning process.
	\item Ordinary differential equations (ODEs) are typically solved for a fixed set of initial conditions, and changes in those conditions require re-solving the system. PINNs, however, can be designed to handle varying initial conditions without depending on previous solutions \cite{flamant2020solving}. In general, boundary and initial conditions are incorporated into the loss function in PINNs, which makes handling complex conditions significantly easier in challenging cases.
	\item PINNs can leverage knowledge from related problems. For instance, through transfer learning, a PINN trained on one problem can be efficiently adapted to solve a similar one \cite{desai2021one}.
	\item Classical approaches struggle with noisy or incomplete data, a common occurrence in experimental settings \cite{karniadakis2021physics}. PINNs, however, can incorporate such data more effectively.
	\item Popular solvers such as the Finite Element Method (FEM) or Finite Difference Method (FDM) often face challenges with mesh generation. PINNs avoid this issue, as they rely only on collocation points, making them inherently mesh-free \cite{karniadakis2021physics}.
	\item Solving inverse problems using traditional numerical approaches can be computationally prohibitive \cite{karniadakis2021physics}, whereas PINNs provide a more efficient framework for such problems.
\end{itemize}
On the downside: 
\begin{itemize}
	\item Training PINNs can be slow, and convergence issues often arise.
	\item Vanishing and exploding gradients pose significant challenges during training.
	\item Balancing the weights assigned to different loss terms is difficult \cite{Yu2022}.
	\item Owing to their iterative nature, PINNs may achieve lower accuracy than classical methods and often require long training times.
	\item Although the curse of dimensionality is less severe in PINNs compared to classical methods, it still presents challenges. Of particular relevance to structural mechanics is that nonlinear higher-order PDEs involve derivatives of varying orders, which significantly affect the accuracy of automatic differentiation and may lead to convergence issues \cite{he2024}. This problem arises even in computing second-order derivatives \cite{sirignano2018dgm} and in classical local beam mechanics governed by fourth-order ODEs. In the present nonlocal case, the governing sixth-order ODE poses an even greater challenge. For these reasons, higher-dimensional problems become especially problematic when higher-order derivatives are involved.
\end{itemize}	

One of the first attempts to apply PINNs to local static beam problems is \cite{taniya}, in which forward and inverse complex beam problems were analyzed. Both Bernoulli-Euler and Timoshenko beam theories were addressed. It was found that PINNs can solve complex beam systems in an efficient and robust manner, even in the presence of noisy data. However, the authors pointed out the necessity of reducing computational cost and issues related to generalizability. In \cite{BAZMARA2023152}, PINNs were employed to model nonlinear bending of 3D functionally graded beams, while in \cite{bazmara2023application}, a similar framework was applied to analyze buckling of the same class of beams, now resting on a Winkler-Pasternak foundation. Bending of porous cantilever microbeams were studied in \cite{tariq2025bending} using PINNs and the modified couple stress theory, including size effects and employing Bernoulli-Euler beam theory. Nonlocal bending was also investigated in \cite{mirsadeghi2025physics} using the nonlocal strain gradient theory. A key aspect of this work is the systematic optimization of hyperparameters via Gaussian process-based Bayesian optimization. While this approach is feasible for computationally less expensive problems, such as static bending, its repetitive nature can make it almost impossible to apply to the computationally much more demanding eigenvalue problems. Another application of PINNs to nonlocal bending of beams, using Eringen's theory, was presented in \cite{kianian2025pinn}. The influence of a three-parameter nonlinear elastic foundation was investigated, and both pure forward and pure inverse problems were considered. It was found that increasing the number of collocation points has a negligible effect on computational cost. Furthermore, the inverse solution did not depend on the starting point, and the PINN performed satisfactorily in the presence of noise. Finally, the study concluded that using an equispaced distribution of collocation points is preferable to a random distribution.	

PINNs can also be applied to solving eigenproblems. In this context, they are regarded as an unsupervised forward-inverse technique, where the forward component models the governing differential equation and the inverse component estimates the corresponding eigenvalue. The literature in this area is still relatively limited. For example, \cite{jin2020unsupervised} proposed an unsupervised PINN algorithm to solve eigenproblems such as the infinite square well and the quantum harmonic oscillator, both governed by the second-order  Schr{\"o}dinger equation. This method was later improved by incorporating a normalization condition in \cite{Jin2022}. More recently, \cite{HARCOMBE2023102136} investigated the  Schr{\"o}dinger equation in the context of discovering localized eigenstates. A notable earlier contribution was made in \cite{LAGARIS19971}, who proposed an ANN framework for solving eigenvalue problems in quantum mechanics. To identify higher-order eigenstates, additional localization loss terms and orthogonality conditions were introduced into the loss function. Another study \cite{grubisic2021deep}, employed physics-informed dense deep networks and constitutional neural networks to examine eigenmode localization in a class of elliptic reaction-diffusion operators. Furthermore, \cite{vibration5020020} developed and optimized a deep Gaussian process surrogate model to predict the stability behavior of friction-induced vibration under variability, enabling complex eigenvalue analysis.

For the purposes of the present research, PINN-based eigenanalyses in structural mechanics are particularly relevant. A novel PINN algorithm proposed in \cite{Yoo2025} computes the natural frequencies of a cantilever beam based on classical mechanics, while addressing convergence issues and the tuning of associated hyperparameters. Similarly, \cite{Fallah2024_3d} developed a PINN model to capture higher natural frequencies for the free vibration of a three-dimensional functionally graded porous local beam resting on an elastic foundation, obtaining benchmark results consistent with analytical solutions. More precisely, in \cite{Fallah2024_3d}, the exact analytical solution of the differential equation for the mode shape is used to avoid normalization and simplify the calculations. Such an approach belongs to the class of supervised PINNs, which combine classical data-driven learning with physics-based constraints. This represents a significantly simplified framework compared to the unsupervised learning used in \cite{Yoo2025, Jin2022, HARCOMBE2023102136}, where the solution is assumed to be unknown and no labeled data is provided. Instead, solution is learned naturally through the training of the PINN using only the governing differential equation and the boundary and initial conditions. Moreover, since the solution is assumed to be known in \cite{Fallah2024_3d}, this can be classified as a sort of an inverse problem, whereas the case in which both the eigenmode and the eigenvalue are unknown belongs to the more complex forward–inverse category. The unsupervised forward-inverse approach is adopted in the present research as well. However, to the best of the authors' knowledge, no studies have yet addressed eigenanalysis for beams exhibiting nonlocal behavior.

The main contributions of the present research can be now summarized as follows:
\begin{itemize}
	\item For the first time a PINN algorithm for performing eigenanalysis of nonlocal beams is presented. This requires solution of a sixth-order problem compared to the local problem that is governed by the fourth-order problem. This also means that, in addition to the standard boundary conditions used in local problems, the constitutive boundary conditions have to be included. All loss terms are physically justified.
	\item The eigenvalue problem analyzed here belongs to the class of unsupervised learning, in which no labeled data is used during learning. Earlier works of this type were limited to the Schr{\"o}dinger equation, i.e. second-order problems \cite{Jin2022, HARCOMBE2023102136} or local beams, i.e. fourth-order problems \cite{Yoo2025}. Hyperparameters used in these works are revisited and modified to handle sixth-order problems. For example, the point-wise normalization loss is replaced with a normalization loss enforced over the entire domain. This is also different to \cite{Fallah2024_3d}, which uses supervised learning in the fourth-order eigenvalue problem involving beams.
	\item An analysis of the influence of key hyperparameters and various extensions of the training procedure on the solution is presented. More precisely, the adaptive selection of weights, the influence of initial points, batching and random selection of collocation points, different types of normalization constraints, various activation functions, and regularization techniques were investigated. A preliminary application to higher modes is also presented.
\end{itemize}

\section{Stress-Driven Nonlocal Elasticity for Nonlocal Beams}
\label{2}
Eringen's nonlocal theory of elasticity assumes that the stress at a point depends not only on the elastic strain at that point but also on the elastic strains at all other points in the structural domain, weighted by an averaging kernel function. However, as mentioned earlier, this theory has shown ill-posedness \cite{romano2017constitutive} in structural mechanics problems, showing incompatibility between equilibrium and constitutive requirements. The stress-driven approach on the other hand, circumvents such issues by defining the elastic strain as a nonlocal function of stresses, ensuring well-posedness for a broader range of boundary value problems, especially in nanomechanics, and it will be thus exploited in the following.

A slender Bernoulli–Euler beam of length $L$ and mass per unit length $m$ is analyzed under bending, with $K_b$ denoting the local elastic bending stiffness, and $M$ the bending moment. The bending occurs in the Cartesian $(x, y)$-plane, where the $ x $-axis aligns with the beam's longitudinal axis. 
The field $ w:[0, L] \to \mathcal{R} $ describes the beam's transverse displacement. The linearized geometric curvature field \( \chi:[0, L] \mapsto \mathcal{R} \) kinematically compatible with the transverse displacement is
\begin{equation}
    \chi \,  =\partial_x^2 w= \, \chi^\mathrm{el} \, + \, \chi^\mathrm{nel} \,\,, 
       \label{elasticcurvature}
\end{equation}
where the total curvature field  $\chi$ is given by the sum of elastic curvature field $\chi^\mathrm{el}$ and nonelastic curvature field $\chi^\mathrm{nel}$. 
In this study, a stress-driven nonlocal integral formulation is exploited to study the dynamics of a cantilever nanobeam. According to the stress-driven integral model of elasticity of slender beams, the elastic flexural curvature field $\chi^\mathrm{el}$ is the convolution of the local source field $\dfrac{M}{K_b}\,$ and a proper averaging kernel $\phi_{a}$ described by the characteristic length ${l_b}\,$:
\begin{equation}
    \,\chi^\mathrm{el}= \int_0^L \phi_{a}\,(x,\Bar{x})\ \frac{M}{K_b}\,(\Bar{x})\,d\Bar{x}\, .
    \label{stress-driven}
\end{equation}

The characteristic length ${l_b}\,$ in \eqref{stress-driven} is defined by $l_b:= a L$ where $a$ is a nonlocal parameter.It is worth noting that the beam characteristic length ${l_b}$ describes the extent of long-range interaction forces in a body and it is closely related to the interatomic distance \cite{Eringen19834703}.
The averaging kernel is chosen as the bi-exponential function 
\begin{equation}
    \phi_{a}(x)=\frac{1}{2 \, l_b} e^{-\frac{|x|}{l_b}} \,,
    \label{Helmholtzkernel}
\end{equation}
that fulfills symmetry, positivity and limit impulsivity as:
{\begin{equation}
   \begin{cases}
        \phi_{a}(x - \bar{x}) = \phi_{a}(\bar{x} - x) \geq 0 \, ,\\
        
        \int_{-\infty}^{+\infty} \phi_{a}(x) dx = 1 \, , \\
        \lim_{a \to 0^+} \int_{-\infty}^{+\infty} \phi_{a}(x - \bar{x}) f(\bar{x}) d\bar{x} = f(x)  ,
    \end{cases}
    \label{kernelprop}
\end{equation}
for any continuous function \( f: \mathbb{R} \rightarrow \mathbb{R} \).
Alternative kernels such as bi-Helmholtz averaging kernels (as adopted in \cite{Zhang2022Nonlocal,Zhang2025Truth}) can be also exploited.

Thanks to the properties fulfilled by the special kernel in Eq.~\eqref{Helmholtzkernel}, the spatial convolution in Eq.~\eqref{stress-driven} can be efficiently inverted providing the following differential equation \cite{romano2017constitutive}:
\begin{equation}
    \chi(x)- l_b^2 \, \partial_x^2 \chi(x)=\left( \frac{M}{K_b} \right) (x) \, ,
    \label{diffeqbeam}
\end{equation}

\noindent equipped with the following constitutive boundary conditions
\begin{equation}
    \begin{cases}
        &\partial_x \chi (0)-\dfrac{1}{l_b}\chi (0)=0 \, , \\
        &\\
        &\partial_x \chi (L)+\dfrac{1}{l_b}\chi (L)=0 \, .
    \end{cases}
    \label{constitutivebcs}
\end{equation}

In Eq.~\eqref{diffeqbeam}, absence of non-elastic effects has been considered, so that the elastic curvature $\chi^\mathrm{el}(x)$ is equal to the total curvature field.

\section{Dynamic Formulation of Nonlocal Beams}
\label{3}

In this section, the free vibration of the nonlocal beam has been addressed with its respective governing differential equation. An Bernoulli-Euler beam is considered having mass linear density defined along the beam axis. As we are considering the free vibration of the nonlocal beam and there is no external loading applied to the system \cite{romano2017stress}, we can write the equilibrium equation for the beam by applying d'Alembert's principle as:
\begin{equation}
    \partial_x^2 M(x,t) =  -m\partial_t^2 w(x,t) \,.
\end{equation}

 By combining equilibrium with constitutive and kinematic compatibility conditions and assuming spatial uniformity of local bending stiffness $K_b$, the following set of differential equations is obtained, id est
\begin{equation}
    \begin{cases}
        &\partial_x^2 M(x,t) = - m \, \partial_t^2 w(x,t) \, ,\\
        & \\
        &K_b \, \left( \chi(x,t)- l_b^2 \, \partial_x^2 \chi(x,t) \right)= M(x,t) \, ,\\
        & \\
        &\chi(x,t)= \partial_x^2 w(x,t) \,.
    \end{cases}
\end{equation}

This leads to the sixth-order differential equation:
\begin{equation}
    l_b^2 \,  \, \partial_x^6 w(x,t) - \, \partial_x^4 w(x,t) +\frac{m}{K_b} \, \partial_t^2 w(x,t) = 0 \, .
    \label{governingeq}
\end{equation}

To generalize the calculation by removing specific values of the physical quantities involved and to clarify it from a computational perspective, the following dimensionless parameters are used
\begin{equation}
\label{nondim}
    \Bar{x}=\frac{x}{L}, \,    \lambda=\frac{m \, \omega^2}{K_b}L^4 ,\,   a=\frac{l_b}{L} ,\,\Bar{w}=\frac{w}{L}, \Bar{M}= \frac{ML}{K_b}\,.
\end{equation}

Therefore, by exploiting positions in Eq.~\eqref{nondim}, Eq.~\eqref{governingeq} can be rewritten in the following non-dimensional form

\begin{equation}
    a^2 \,  \, \partial_{\Bar{x}}^6 \Bar{w}(\Bar{x},t) - \, \partial_{\Bar{x}}^4  \Bar{w}(\Bar{x},t) +\lambda \,  \Bar{w}(\Bar{x},t) = 0 \, ,
    % l_b^2 \,  \, \partial_x^6 w(x,t) - \, \partial_x^4 w(x,t) +\lambda \, w(x,t) = 0 \, ,
    \label{eigenequation}
\end{equation}
where $\lambda $ is the respective eigenvalue from which  the natural frequency  $\omega$ of the system could be obtained. The derivation of Eq.~\eqref{eigenequation} is proposed in \cite{Vaccaro2021}. Note that, for the sake of notational economy, the overbar will be omitted in the remainder of the text, with the understanding that all quantities retain their nondimensional character.

The subsequent section describes the solution of this sixth-order eigenvalue problem with the help of Physics-Informed Neural Network. 

\section{Physics-Informed Neural Network for Nonlocal Beams}
Physics-Informed Neural Networks have shown efficiency in solving differential equations \cite{RAISSI2019686}. They are a class of machine learning models that incorporate physical laws, typically expressed as differential equations, into the neural network training process. PINNs leverage neural networks to approximate solutions to differential equations while enforcing physical principles (e.g., conservation laws, boundary conditions) as constraints. They combine observational data with the governing equations to ensure that the resulting solutions respect both the data and the underlying physics.

\subsection{PINN Methodology}
\label{sec_pinn_met}
A general overview of the PINN methodology has already been presented in several studies \cite{RAISSI2019686, YUAN2022111260, Fallah2024_3d}. Therefore, this section outlines the specific strategy for solving eigenvalue problems using the PINN framework.

A typical PINN is a feedforward neural network with multiple hidden layers and suitable activation functions. The inputs are the independent variables (e.g., spatial coordinates $x$), and the outputs represent the dependent variables (e.g., displacement $w(x))$. The network is trained to minimize a loss function that includes contributions from the PDE residual, boundary conditions, and any additional regularization terms. Once the relevant loss components are defined, the neural network is trained using an appropriate optimizer (e.g., the Adam optimizer in this study). During training, the network adjusts the weights and biases of its neurons to minimize the total loss. This minimization process leads to accurate predictions of the underlying physical problem. Automatic differentiation is employed to compute derivatives of the network output, enabling the evaluation of the differential operators required by the governing PDE.

Let us consider a general form of an eigenvalue problem as:
 \begin{equation}
\mathcal{L} w(x) + \lambda w(x)= 0
\label{generaleigenform}
\end{equation}
where \( \mathcal{L} \) is a differential operator, \( w \) is the eigenfunction, and \( \lambda \) is the associated eigenvalue.
After feeding on the input into suitable hidden layers using relevant activation function, losses are computed. As we are interested in computing eigenvalues and corresponding eigenfunctions of the system, the total loss is typically defined \cite{Yoo2025} as:

\begin{equation}
L_{\text{tot}} = \alpha_{\mathrm{de}} L_{\mathrm{de}} +  \alpha_{\mathrm{bc}} L_{\mathrm{bc}} +  \alpha_{\mathrm{cbc}} L_{\mathrm{cbc}} + \alpha_{\mathrm{nor}} L_{\mathrm{nor}},
\label{totalloss}
\end{equation}
where \( L_{\mathrm{de}} \) is the differential equation loss, \( L_{\mathrm{bc}} \) and \( L_{\mathrm{cbc}} \) enforce the boundary conditions and constitutive boundary conditions, and \( L_{\mathrm{nor}} \) is the normalization or regularization term. Note that in the present nonlocal case the term $L_{\mathrm{cbc}}$ must be introduced, while in the local case \cite{Yoo2025} this is not required. These individual losses are multiplied by corresponding weighting factors $\alpha_\bullet$ in order to better emphasize each of the above terms.

An important point regarding the definition the normalization should be noted. \cite{jin2020unsupervised} initially proposed a regularization loss term which was given as the summation of losses as shown in Eq.~\eqref{regloss} below, scaled by respective weight factor. The two losses of interest with the role to penalize convergence to trivial solutions were:
\begin{equation}
\label{regloss}
L_\mathrm{nor}= L_\mathrm{f} + L_\lambda, \quad L_\mathrm{f} = w(x)^{-2}, \quad L_\lambda = \lambda^{-2},
\end{equation}
where \( w \) is the predicted eigenfunction, and \( \lambda \) is the eigenvalue. Therefore, we initially adopted Eq.~\eqref{regloss} but PINN was not being able to converge to the correct eigenstate. Investigating on this context, \cite{Jin2022} noted that while the numerical trick in Eq.~\eqref{regloss} tends to penalize trivial solutions, it lacks a physical basis. Moreover, small eigenvalues and eigenfunction close to $w(x)=0$ lead toward infinite values of the loss function, and consequently cause convergence issues. This formulation was updated in \cite{Jin2022, HARCOMBE2023102136} with additional conditions that address these issues. The regularization term is thus proposed in the form of a normalization loss:
\begin{equation}
L_{\text{nor}} = \left( \int_{0}^{L} w^2(x) \, dx - 1 \right)^2,
\label{normloss}
\end{equation}
where \( w \) is the predicted eigenfunction, and the integral is taken over the interval \([0, L=1]\). Application of the midpoint rule provides an approximation as:
\begin{equation}
	L_{\mathrm{nor}} \approx \left(\sum_{i=1}^{N_\mathrm{c}-1} (w_m)_i^2  - \frac{1}{\Delta x} \right)^2,
	\label{normloss_approx}
\end{equation}
where $dx \approx \Delta x = (N_\mathrm{c}-1)/L$, $N_\mathrm{c}$ is the number of collocation points, and $(w_m(x_m))_i$ are the values of the eigenfunction $w(x)$ evaluated at the midpoints $(x_m)_i$ between two neighboring collocation points. We adopted the same approach, which significantly reduced the numerical error and enabling the convergence of the loss term to zero, enabling accurate prediction of the eigenvalue and eigenfunction.
	
Note that although $L_{\mathrm{nor}}$ as defined by Eq.~(\ref{normloss_approx}) and used in \cite{Jin2022} serves the same purpose as the alternative form $L_{\mathrm{nor}} \approx \left(\sum_{i=1}^{N_\mathrm{c}-1} (w_m)_i^2\Delta x - 1 \right)^2$ employed in \cite{HARCOMBE2023102136}, their weighting factors cannot be identical. This follows from the relation $\left(\sum_{i=1}^{N_\mathrm{c}-1} (w_m)_i^2\Delta x  - 1 \right)^2=\Delta x^2\left(\sum_{i=1}^{N_\mathrm{c}-1} (w_m)_i^2  - (\Delta x)^{-1} \right)^2$. In particular, in \cite{Jin2022} $\alpha_{\mathrm{nor}} = 1$, whereas in \cite{HARCOMBE2023102136} $\alpha_{\mathrm{nor}} = (N_\mathrm{c} - 1)^3$ (with $\alpha_{\mathrm{bc}} = (N_\mathrm{c} - 1)^2$ and $\alpha_{\mathrm{de}} = 1$). For the present number of collocation points ($N_\mathrm{c}=100$), this corresponds to a very large weighting factor of approximately $\alpha_{\mathrm{nor}} \approx 10^6$. Finally, it should be noted that \cite{HARCOMBE2023102136} recommends weighting factors that depend on the number of collocation points, thereby increasing the complexity of their determination.

After defining the respective loss terms, the neural network model is trained using a suitable optimizer, preferably the Adam optimizer \cite{Yoo2025,HARCOMBE2023102136,Jin2022}. It is also important to consider the tuning of hyperparameters, such as the number of neurons in dense layers, learning rates, weighting factors for the loss terms, and the choice of activation functions, as these significantly affect the model's training. The optimal adjustment of hyperparameters can vary depending on the problem. However, these aspects are discussed in detail in the case-study sections of this article.

% Section: Case Study
\section{Case Study 1: PINN for Local Cantilever Eigenvalue Problem}
\subsection{Structural Problem Formulation}
The proposed PINN methodology has been first implemented to solve the eigenvalue problem of a local cantilever beam, as illustrated in Fig.~\ref{cantibeam}, to determine the first eigenvalue and its corresponding eigenfunction. Since the local problem is governed by a fourth-order differential equation, its approximation using a PINN requires significantly less computational effort compared to the nonlocal case. Therefore, the selection of the PINN architecture and other relevant hyperparameters is performed on the local problem using a trial-and-error approach, which serves as a basis for subsequent fine-tuning in the nonlocal PINN formulation.

As no specific material has been assigned to the beams, we would like to stick to the non-dimensional eigenform of the differential equation  Eq.~\eqref{eigenequation}. Primarily investigating eigenvalue problem of the local cantilever beam Eq.~\eqref{eigenequation} can be squeezed to the following equation by putting the nonlocal parameter $a=0$

% All beams analyzed in this study have a unit length and a unit square cross-section, with length, width, and height each set to unity. Since no specific materials are assigned to the structural schemes, the combination of geometric and material parameters, $\frac{m}{K_b}$, is also taken as 1. In the local case, the governing equation is a fourth-order differential equation, as shown in Eq.~\eqref{eigenequation} with $l_b=0$:
\begin{equation}
    - \partial_{{x}}^4 {w} + \lambda {w} = 0, \quad {x} \in [0, 1],
\end{equation}
and the beam is subjected to its respective kinematic boundary conditions at ${x}=0$ and at ${x}=1$:
% Listing boundary conditions
\begin{itemize}
	\item {Clamped at ${x}=0$}:
	\begin{equation}
		{w}(0) = 0, \partial_{{x}} {w}(0) =0.
		\label{clampedbcLOC}
	\end{equation}
	\item {Free at ${x}=1$}:
	\begin{equation}
		{M}(1)=\partial_{{x}}^2{w}(1)= 0, \partial_{{x}} 
        {M}(1)=\partial_{{x}}^3 {w}(1)= 0.
		\label{freebcLOC}
	\end{equation}
\end{itemize}
The PINN approach is implemented to approximate the eigenfunction $w$ and the eigenvalue $\lambda$, that correspond to the square of the natural frequency $\omega$. 

\begin{figure}
    \centering
    \includegraphics[width=0.7\linewidth]{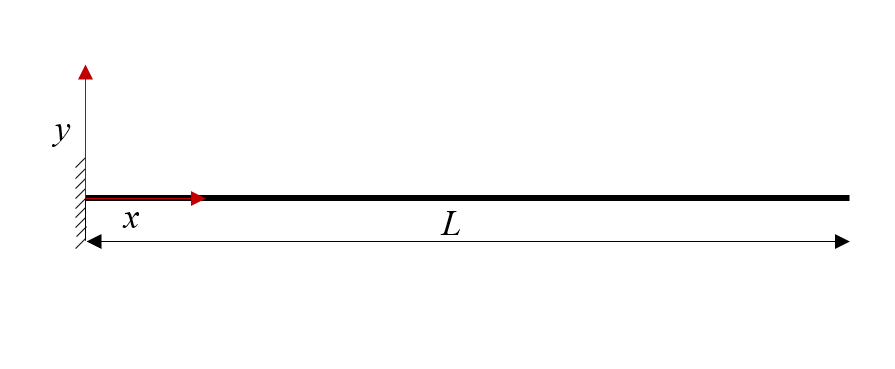}
    \caption{Schematic of Nonlocal Cantilever Beam}
    \label{cantibeam}
\end{figure}

% Subsection: Step-by-Step Implementation
\subsection{PINN Architecture and Training}
\label{sec_PINNlocal}
In all cases considered in the present research, we adopted a PINN model, as shown in Fig.~\ref{PINN}, consisting of two components: a neural network to approximate the eigenfunction $w$ and a trainable variable representing the eigenvalue $\lambda$. An L2 regularization penalty of $10^{-6}$ was applied. The neural network consists of three dense layers with 128, 64, and 32 neurons, respectively, using the swish activation function
\begin{equation}
    f(x)=x\cdot\sigma(\beta x), \quad \mathrm{where} \quad \sigma(\beta x)=\left(1+\exp(-\beta x)\right)^{-1}, \; \beta=1
\end{equation}
in the hidden layers, \cite{ramachandran2017searching}. While activation functions such as hyperbolic tangent, ReLU, softplus, sine, and sigmoid are commonly used in PINNs, we observed that the swish activation function significantly improved the accuracy of eigenstate prediction. This finding is consistent with reports in the literature \cite{al2021time}, which attribute the improvement to swish's ability to mitigate the vanishing gradient problem. Although the ReLU activation function was successfully applied to the second-order problem in \cite{HARCOMBE2023102136}, it is not expected to be a suitable choice for the sixth-order problem at hand due to the absence of higher-order derivatives. The hyperbolic tangent activation function has been used in the classical cantilever beam eigenvalue fourth-order problem. The sine function was employed in \cite{jin2020unsupervised,Jin2022}, where superior performance compared to the hyperbolic tangent and sigmoid functions was reported. Bias vectors were not used in the network, and all layers employed the Glorot uniform initializer for weight initialization. In the initial phase of the research, random seeds were fixed in TensorFlow to ensure reproducibility; in later stages, random seeds were not fixed, but the performance remained consistent. 

The Adam optimizer was employed with a relatively large constant learning rate of $10^{-2}$. Note that while a high learning rate (as used in the present section) may accelerate convergence during the early stages of training, it can also cause oscillations or even divergence in the loss functions at later stages, as the optimizer may overshoot the minima. Conversely, using a very low learning rate (see \ref{sec:app1}) can result in slow convergence and may cause the optimizer to become trapped in local, rather than global, minima. Unfortunately, there is no universal method for determining the optimal learning rate other than through a trial-and-error procedure, which is also the case here. This is completely in line with \cite{RAISSI2019686} and subsequent works. The choice of learning rate depends on multiple factors, including problem stiffness, the weighting of loss terms, network depth, activation functions, optimizer type, normalization, manual or adaptive loss weighting, etc. Frequently, two-stage procedure is often effective, using a high learning rate at the beginning, and after some criterion is met, the learning rate is reduced near the minima to increase the accuracy of the solution. Automatic learning rate decay can be also successfully employed. In this way, the monolithic scheme employing both higher and lower learning rates, as used in the present manuscript, can be extended into a two-stage training procedure.

\begin{figure} [h!]
    \centering
    \includegraphics[width=0.8\linewidth]{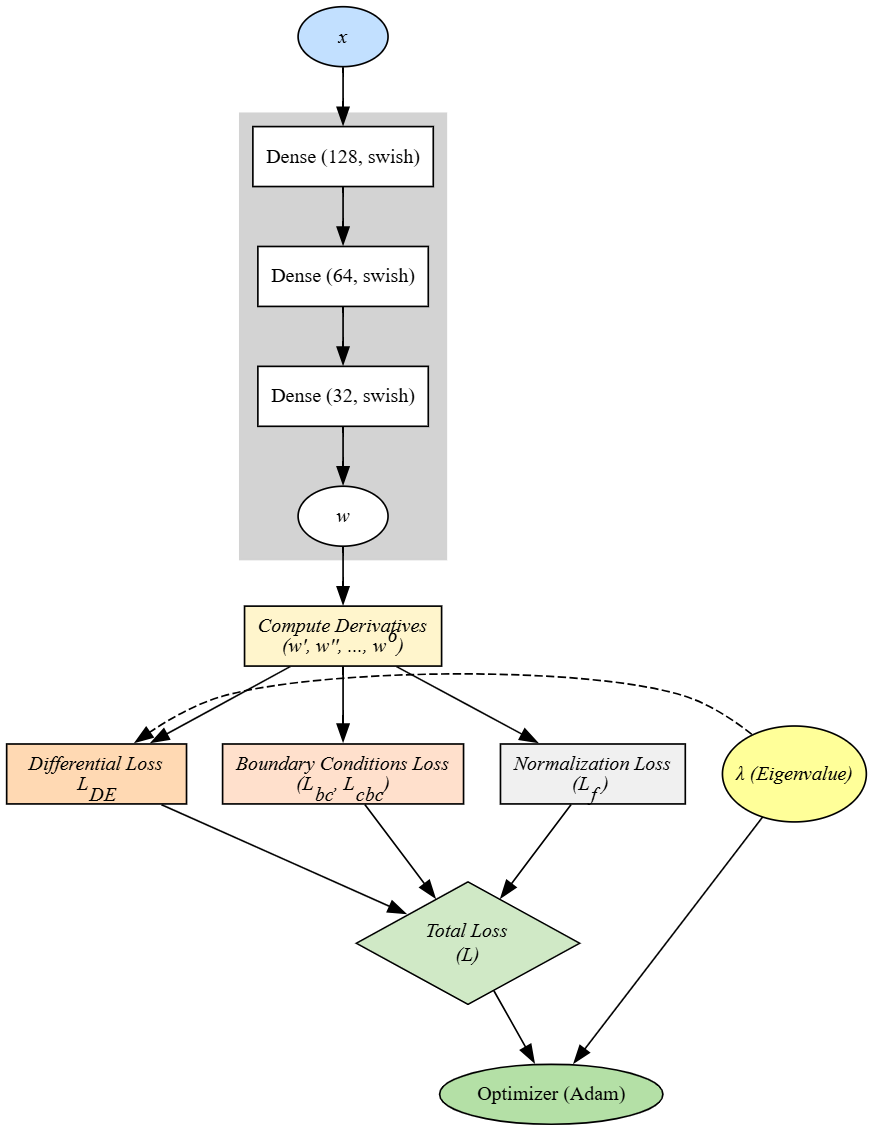}
    \caption{PINN Architecture}
    \label{PINN}
\end{figure}

A specific initial value of $\lambda$ is provided, which gradually adjusts itself during training to attain the desired eigenvalue. We generate 100 uniformly spaced collocation points in ${x}\in [0, 1]$ for the cantilever beam to evaluate the differential equation residual and boundary conditions. Note that it is sometimes advantageous to use randomly distributed collocation points in order to improve training in PINNs by enhancing sampling diversity or mitigating overfitting. However, in the present unsupervised eigenvalue problem, which involves a smooth beam and high-order spatial derivatives, this strategy does not offer any advantage. The eigenmodes of a beam are smooth and well-behaved functions, and the governing equation is linear with constant coefficients. Using random collocation points in such a case may lead to non-uniform sampling density and numerical instability when evaluating high-order derivatives through automatic differentiation. Similar conclusions were reported in \cite{kianian2025pinn}. However, for completeness, an analysis of batching and random selection of collocation points is provided in \ref{sec:app5_batches}. No particular advantage of random selection of collocation points was observed. Therefore, as already stated, uniformly spaced collocation points were adopted in this work. This approach ensures consistent numerical accuracy, stable evaluation of derivatives up to the fourth or sixth order, and a uniform error distribution across the domain, all of which are crucial for achieving reliable convergence in smooth, high-order eigenvalue problems.

The model is trained for a given number of epochs, updating weights and $\lambda$ at each epoch. The training loop computes fourth-order derivatives using automatic differentiation via \texttt{tf.GradientTape}, and evaluates the loss components $L_{\text{de}}$, $L_{\text{bc}}$, and $L_\text{nor}$. The weighting factors for the losses were set as $\alpha_{\text{de}}=1$, $\alpha_{\text{bc}}=100$, and $\alpha_\text{nor}=5$. Selecting such high values for the boundary condition weight helps stabilize the solution. Note that in the case of the local cantilever eigenproblem in \cite{Yoo2025}, the same weights, $\alpha_{\text{de}} = 1$ and $\alpha_{\text{bc}} = 100$, were used. However, instead of the normalization constraint in Eq.~(\ref{normloss}), which is imposed over the entire domain, a constraint in \cite{Yoo2025} was applied at a single point, namely the cantilever tip, $w(1) = 1$, see \ref{sec_yoo}. The latter constraint is clearly weaker, since it enforces normalization only at a single point, whereas the physical condition in Eq.~(\ref{normloss}) could still be violated. Also, \ref{sec_yoo} shows that normalization at a single point is somewhat worse than normalization over the entire domain. Nevertheless, the method of \cite{Yoo2025} has one notable advantage. The most logical choice for the weight of the normalization condition is to take it the same as that for the boundary conditions. This is particularly helpful in the analysis of higher modes, which may otherwise require adjusting of the normalization weight used in the present research. During the adjustment of weighting factors by trial-and-error, it was observed that larger weights for boundary conditions facilitate training. The scaling factor $\alpha_{\text{de}}$ applied to the loss is kept at 1, signifying that this loss component is neither amplified nor diminished relative to the other components in the total loss history. Model weights and training state are saved at each new minimum of the total loss. After training, the final eigenfunction ${w}$ is normalized such that ${w}(1) = 1$. For an attempt to use adaptive weights, see \ref{sec:app4_adaptive}. It is found that manually selected weights yields better results.

Finally, it is well known that the use of batches often improves the training efficiency of neural networks. This approach avoids computing gradients over the entire dataset at once, relying instead on approximate gradients obtained from batches, which can accelerate training in conventional machine learning problems. However, in the present type of PINN, the situation is different. Each batch requires recomputation of high-order derivatives (4$^\mathrm{th}$ or 6$^\mathrm{th}$ order in this case) for all collocation points within the batch. This is computationally very expensive due to multiple evaluations instead of a single evaluation in a full-batch case. Since the number of collocation points used here is relatively small ($\approx$100), batching would in fact significantly increase the total computation time rather than reduce it. For these reasons, a full-batch training strategy was adopted in the present work, as it provides the same gradient information at a lower computational cost. For a justification of these claims, see \ref{sec:app5_batches}.

\subsection{Results and Discussion}
All calculations in this research were performed on a DELL XPS 17-9000 PC equipped with 64 GB of RAM and an Intel(R) Core(TM) i7-10750H CPU (2.60 GHz, 6 cores). The computations were carried out using the TensorFlow 2.10.0 library. The main results are presented in Tab.~\ref{tab_local_CF} and Fig.~\ref{fig_local_high_LR}. It is noteworthy that, although the initial value of $\lambda$ differs considerably from the analytically obtained eigenvalue, the PINN is still able to converge to the correct solution. Compared to other authors \cite{Yoo2025, Jin2022, HARCOMBE2023102136}, who used learning rates of $1\times 10^{-3}$, $8 \times 10^{-3}$, and $5 \times 10^{-4}$ for less demanding problems than the current one, the constant high learning rate of $10^{-2}$ employed here - for both $\lambda$ and the network weights - demonstrated robust training performance. It successfully captured the correct first eigenvalue and eigenfunction, albeit at the expense of slightly reduced accuracy. The computational cost for the present local cantilever is about 325 epochs per minute. To further investigate the training behavior under lower and adaptive learning rate, an additional analysis is conducted and documented in \ref{sec:app1}.

\begin{table}
	\begin{center}
	\begin{tabular}{|l|r|}
		\hline
         No. epochs			  & \numprint{30000}    \\
		\hline
		Initial $\lambda$     & 100.0000 \\
		Predicted $\lambda$   & 12.4581  \\
		Analytical $\lambda$  & 12.3596   \\
		Difference (\%) 	  & -0.7972   \\
		\hline
		min. $L_\mathrm{tot}$ & 0.05489   \\
		\hline
	\end{tabular}
	\caption{Number of training epochs, initial, predicted, analytical value of $\lambda$, and difference, total loss for a high learning rate case applied to a local cantilever beam. }
	\label{tab_local_CF}
	\end{center}
\end{table}

\begin{figure}
	\centering
	\includegraphics[width=11cm]{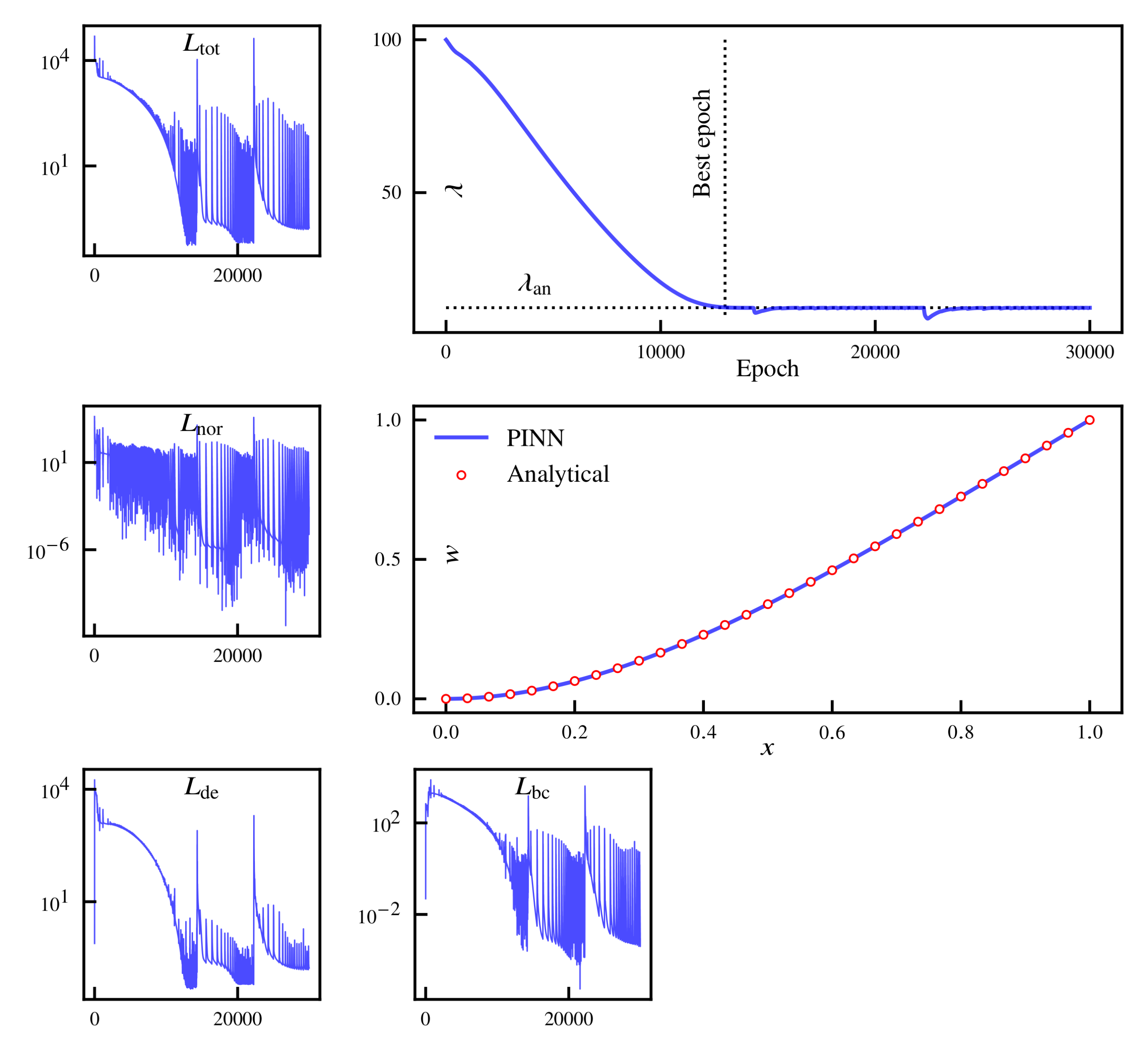}
	\caption{Local cantilever beam with high learning rate: $\lambda$ convergence, first eigenfunction, and loss histories.}
	\label{fig_local_high_LR}
\end{figure}

\section{Case Study 2: PINN for Nonlocal Cantilever Beam Eigenvalue Problem}
\label{CS_nonlocal_canti}
% Subsection: Problem Formulation
\subsection{Structural Problem Formulation}
In this case study, the eigenstate of a nonlocal cantilever beam is considered. The governing equation has already been presented in Eq.~\eqref{eigenequation}, and the nonlocal parameter is taken as $a = 0.1$. 
% Therefore, based on the relation between the characteristic length and the actual length described above, $l_b$ is also equal to $0.1$ for a unit actual length $L = 1$. 
The boundary conditions for the nonlocal cantilever beam include both the constitutive boundary conditions, as defined in Eq.~\eqref{constitutivebcs}, and the classical boundary conditions. These can be listed as:

% Listing boundary conditions
\begin{itemize}
	\item {Clamped at ${x}=0$}:
	\begin{equation}
		\begin{array}{c}
		{w}(0) = 0 \\
		\partial_{{x}} {w}(0) = 0 \\
		 a\partial_{{x}}^3 {w}(0) - \partial_{{x}}^2 {w}(0) = 0.
		\end{array}
		\label{clampedbc}
	\end{equation}
	\item {Free at ${x}=1$}:
	\begin{equation}
		\begin{array}{c}
		{M}(1)= \partial_{{x}}^2 w(L) - a^2 \partial_{{x}}^4 {w}(1) = 0 \\
		\partial_{{x}}{M}(1)= \partial_{{x}}^3 {w}(1) - a^2 \partial_{{x}}^5 {w}(1) = 0 \\
		 a\partial_{{x}}^3w(1) + \partial_{{x}^2} {w}(1) = 0.
		\end{array}
		\label{freebc}
	\end{equation}
\end{itemize}

It should also be noted that the values of $\lambda$ correspond to the eigenvalues of the problem, and the natural frequency $\omega$ can be computed as  square root of $\lambda$, since the other material and geometric parameters are taken as unity in the present case.

\subsection{Design of Loss Functions}
As mentioned in Sec.~\ref{sec_pinn_met}, appropriately defining the loss functions plays a crucial role in training the neural network to solve differential equations by embedding physical laws and constraints directly into the optimization process. Below is a concise explanation of the different types of loss functions used in our PINN model, including the differential equation loss $L_{\text{de}}$, kinematic boundary loss $L_{\text{bc}}$, constitutive boundary loss $L_{\text{cbc}}$, and normalization loss $L_{\text{nor}}$.

\begin{itemize}
	\item Differential Equation Loss ($L_{\text{de}}$):
	\begin{equation}
		L_{\text{de}} = \left[ \frac{1}{N_\mathrm{c}} \sum_{i=1}^{N_\mathrm{c}} \left( a^2 \partial_{{x}}^6 {w}({x}_i) - \partial_{{x}}^4 {w}({x}_i) + \lambda {w}({x}_i) \right)^2 \right]
		\label{DEloss}
	\end{equation}
The loss $L_{\text{de}}$ represents the squared residual of the differential equation evaluated at each collocation point ${x}_i$. The residual quantifies how well a candidate solution $w({x})$ satisfies the differential equation. The factor $\frac{1}{N_\mathrm{c}}$ normalizes the sum of squared residuals by the number of collocation points, ensuring that the loss represents an average and is independent of the discretization density. This normalization allows for consistent comparisons across different sets of collocation points. The summation $\sum_{i=1}^{N_\mathrm{c}}$ aggregates the squared residuals at all collocation points, with the index $i$ running from 1 to $N_\mathrm{c}$, so that the loss evaluates the differential equation's residual at $N_\mathrm{c}$ distinct points.
	\item Classical Boundary Condition Losses ($L_{\text{bc}}$):
\begin{align}
	L_{\text{bc}} = & {w}(0)^2 + \left( \partial_{{x}} {w}(0) \right)^2 \notag \\
	&+ \left( \partial_{{x}}^2 {w}(1) - a^2 \partial_{{x}}^4 {w}(1)\right)^2 + \left( \partial_{{x}}^3 {w}(1) - a^2 \partial_{{x}}^5 {w}(1) \right)^2  ,
	\label{kinematicbcloss}
\end{align}
The total kinematic boundary condition loss is computed as the mean of the squared terms for all kinematic boundary conditions.

	\item Constitutive Boundary Condition Loss ($L_{\text{cbc}}$):
\begin{equation}
	L_{\text{cbc}} = \left[ \left( a\partial_{{x}}^3 {w}(0) - \partial_{{x}}^2 {w}(0) \right)^2 + \left(  a\partial_{{x}}^3 {w}(1) +   \partial_{{x}}^2 {w}(1) \right)^2 \right],
	\label{customloss}
\end{equation}
	\item Normalization Loss ($L_{\mathrm{nor}}$):
As discussed in Sec.~\ref{sec_pinn_met}, the normalization loss has been defined in Eq.~(\ref{normloss_approx}) as:
\begin{equation}
	L_{\mathrm{nor}} = \left(\sum_{i=1}^{N_\mathrm{c}-1} ({{w}_m})_i^2   - \frac{1}{\Delta {x}} \right)^2.
\end{equation}
During training, it was observed that the scaling factor for the regularization loss plays a crucial role in both the loss and eigenvalue convergence. Therefore, it can be considered an important hyperparameter.
\end{itemize}

The total loss is computed using the weighting factors defined in Eq.~(\ref{totalloss}). The contributions of the individual loss terms and their corresponding weights have been carefully tuned to ensure favorable convergence and accuracy of both the eigenvalue and the eigenfunction. The specific values for these weights are the same as those used in the local case, as described in Sec.~\ref{sec_PINNlocal} with the addition of $\alpha_{\mathrm{cbc}}=100$. Similarly, all other hyperparameters remain unchanged from the local case.

\subsection{Preliminary Investigation}
\label{sec_CS2_preliminary}
In the preliminary stage, the influence of selected hyperparameters, previously determined for the local case, was re-examined for the nonlocal cantilever beam.

The first set of analyses examines the influence of the L2 regularization penalty applied to each dense layer. Gradient clipping on the training variables was not employed. As evident from Tab.~\ref{tab_Nonlocal_CF_L2} and Fig.~\ref{fig_L2}, an L2 penalty of $10^{-6}$ yields the best results for both $\lambda$ and the total loss after training for \numprint{50000} epochs. In contrast, the absence of L2 regularization produces the poorest approximation of the first eigenvalue. Similarly, for the case $10^{-4}$, the best result occurs at epoch \numprint{23322}, after which no further improvement is observed, unlike the other cases. For these reasons, the penalty value $10^{-6}$ is adopted in all subsequent analyses. Finally, it is certainly possible that other L2 values could yield improved results; however, identifying the optimal value would require a more thorough analysis. To keep the number of evaluations within reasonable limits, we simply focused on determining the best option among the three cases investigated.

\begin{table}
	\begin{center}
		\begin{tabular}{|l|r|r|r|}
			\hline
			L2			  		  & None  & $10^{-4}$ & $10^{-6}$   \\
			\hline
			Best epoch			  & \numprint{49578} & \numprint{23322} & \numprint{49774}    \\
		\hline
		Best $\lambda$     	  & 14.8844 & 15.0812 &	15.1236    \\
		\hline
		min. $L_\mathrm{tot}$ & 0.8070 & 1.88343 &  0.5357 \\
		\hline
	\end{tabular}
	\caption{Influence of L2 regularization on PINN training for \numprint{50000} epochs. Analytical result $\lambda_\mathrm{an}=15.1953$, $\lambda_{\mathrm{ini}} = 100$.} 
	\label{tab_Nonlocal_CF_L2}
\end{center}
\end{table}

\begin{figure}
\centering
\includegraphics[width=10cm]{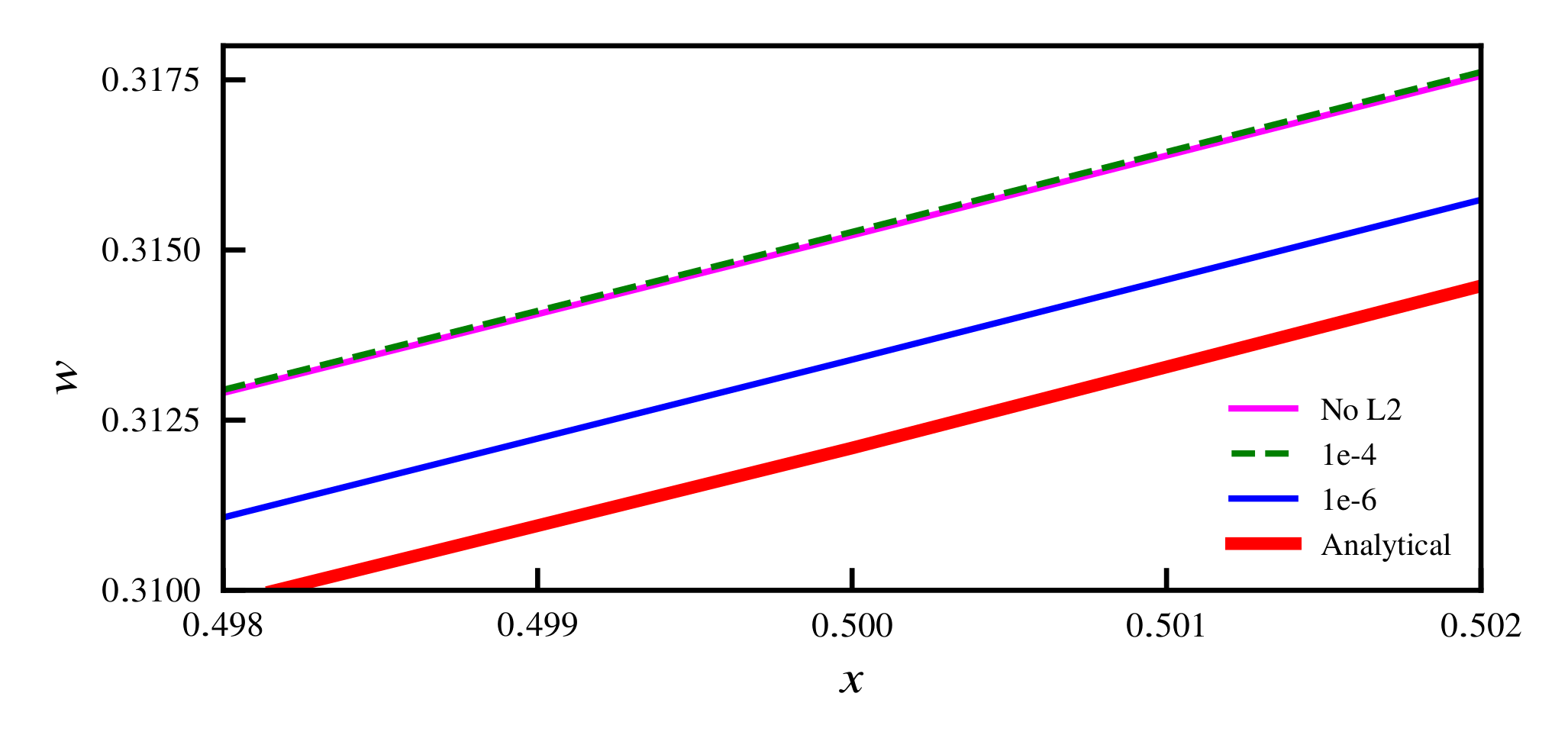}
\caption{Influence of L2 regularization on PINN training - central part of the first eigenfunction curve. Colors: magenta - no L2 regularization, green - regularization value $10^{-4}$, blue - regularization value $10^{-6}$, red - analytical solution.}
\label{fig_L2}
\end{figure}

The second set of analyses focused on the effect of gradient clipping on PINN training. Four cases were considered:
\begin{enumerate}
	\item $\lambda$ gradient clipped by value to remain in the range $\left[-1, 1\right]$, while gradients of the remaining NN weights were clipped by L2-norm at 1.0 (denoted as 1.0 in Fig.~\ref{fig_clip}).
	\item $\lambda$ gradient clipped by value to remain in the range $\left[-1, 1\right]$, while gradients of the remaining NN weights were clipped by L2-norm at 0.1 (denoted as 0.1).
	\item $\lambda$ gradient not clipped, while gradients of the remaining NN weights were clipped by L2-norm at 1.0 (denoted as No clip $\lambda$).
	\item Neither $\lambda$ nor weights gradients were clipped (denoted as No clip).
\end{enumerate}

Part of the results is shown in Fig.~\ref{fig_clip} and Tab.~\ref{tab_Nonlocal_CF_clip}. In all cases, the same random initial weights were used, and an L2 regularization penalty of $10^{-6}$ was applied in each dense layer. It is evident that when gradients were not clipped (No clip, case 4 above), the value of $\lambda$ closest to the analytical solution was obtained, although the total loss was slightly worse than in case 1.0 (case 1 above). However, the most accurate eigenfunction was again observed in case 4 (No clip), Fig.~\ref{fig_clip}. From Tab.~\ref{tab_Nonlocal_CF_clip}, it is also clear that in the No clip $\lambda$ case, the lowest loss is obtained relatively early (epoch \numprint{37958}), after which the optimizer is unable to progress further. A similar conclusion can be drawn for the remaining cases where gradient clipping was applied, although the effect is less pronounced. When gradient clipping is not used (No clip case), the optimizer is able to progress throughout the training and would most likely continue to improve further if more epochs were allowed. Consequently, no gradient clipping was adopted for the main calculations presented in the next section.

\begin{figure}
	\centering
	\includegraphics[width=10cm]{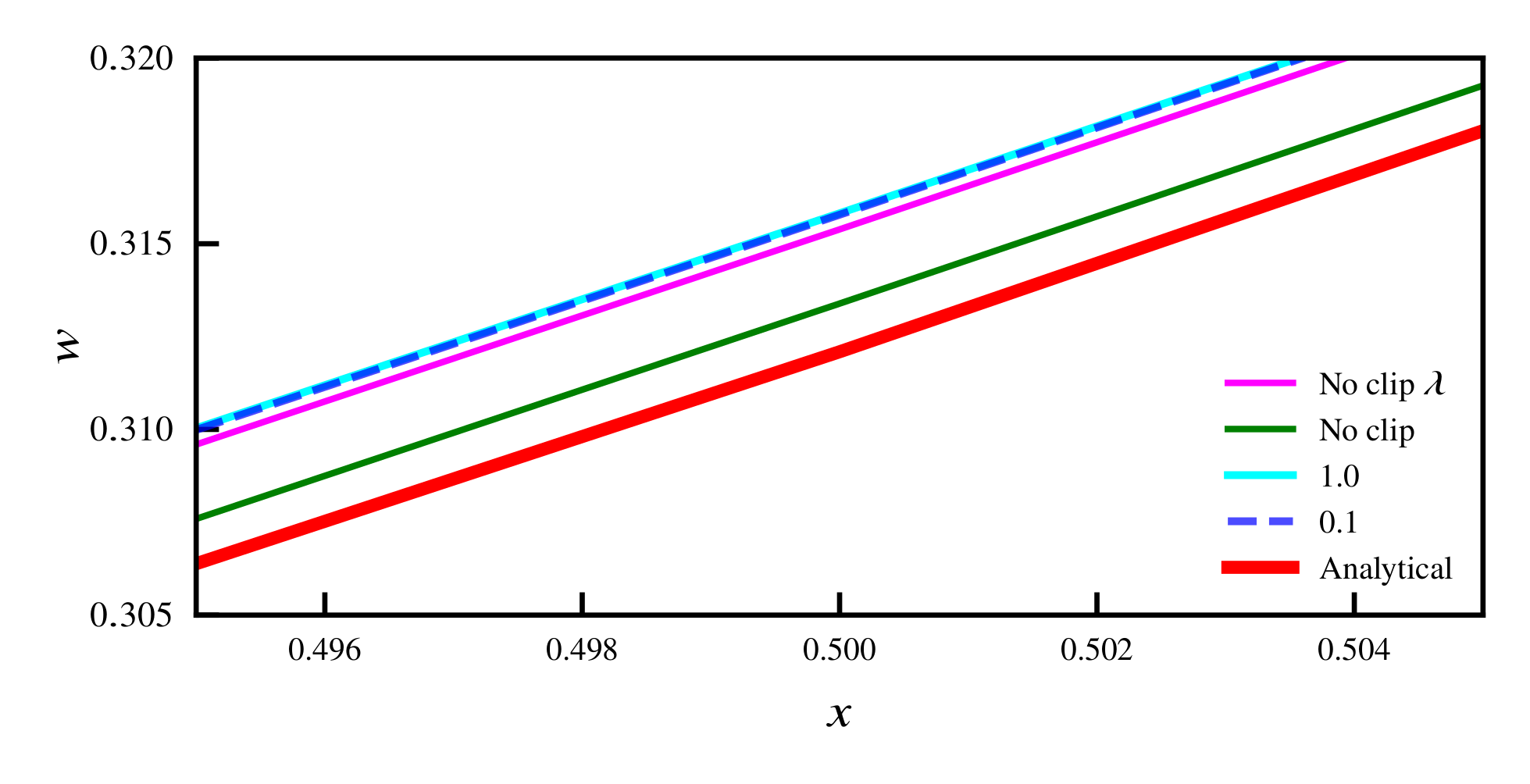}
	\caption{Influence of gradient clipping on PINN training: central part of the first eigenfunction curve. Colors:  green - no gradient clipping, magenta - no gradient clipping for $\lambda$ and NN weights gradients clipping value 1.0, cyan - clipping value 1.0, blue/dashed - clipping value 0.1, red - analytical solution.}
	\label{fig_clip}
\end{figure}

\begin{table}
	\begin{center}
		\begin{tabular}{|l|r|r|r|r|}
			\hline
			Case			  		  & 1.0  & 0.1 & No clip $\lambda$ & No clip   \\
			\hline
			Best epoch			  & \numprint{43800} & \numprint{46635} & \numprint{37958} & \numprint{49774}     \\
			\hline
			Best $\lambda$     	  &  15.049 & 15.037 &	14.869 & 15.124   \\
			\hline
			min. $L_\mathrm{tot}$ & 0.515 & 0.874 &  1.269 & 0.535 \\
			\hline
		\end{tabular}
		\caption{Influence of gradient clipping on PINN training for \numprint{50000} epochs. Analytical result $\lambda_\mathrm{an}=15.1953$, $\lambda_{\mathrm{ini}} = 100$.}
		\label{tab_Nonlocal_CF_clip}
	\end{center}
\end{table}
The third and final set of investigations concerns the choice of activation function. This is an important point, as the choice of activation function can strongly affect the solution accuracy in PINNs. For example, \cite{wang2023learning} analyzed the 1D Poisson equation and reported accuracy values ranging from 0.73\% to 75.77\%, depending on the activation function. Furthermore, the recursive nature of automatic differentiation (i.e., nested derivatives) used to compute $n^\mathrm{th}$-order derivatives can lead to numerical instability and inefficiency \cite{lu2021deepxde}. This issue is particularly critical in forward-inverse problems, where errors in high-order derivatives can significantly affect the solution. Therefore, conclusions drawn from much simpler forward, inverse, or first-order problems, which are less sensitive to such numerical effects, must be applied with caution and reevaluated for the present problem.

As previously mentioned, hyperbolic tangent, ReLU, softplus, sine, sigmoid, and swish activation functions were considered. The convergence of $\lambda$ during training is shown in Fig.~\ref{lambda_act_fun}. It is clear that the swish activation function is by far the most effective choice, while the only viable alternative could be the softplus function, although it exhibits a lower convergence rate. Regarding the sigmoid activation function, more epochs would be needed for a definitive assessment, but preliminary results suggest it may produce a similar outcome, though with markedly slower convergence. Regarding the effect on the first eigenmode, shown in Fig.~\ref{mode_act_fun}, the same conclusions can be drawn. Differently to \cite{HARCOMBE2023102136,Yoo2025, Jin2022, jin2020unsupervised}, hyperbolic tangent, ReLU, and sine functions do not provide an appropriate shape of the curve, although the boundary conditions are correctly satisfied. 

The above behavior can be explained as follows. For the sixth-order differential equation considered here, the network output must be differentiated up to the sixth order. The ReLU activation function lacks second- and higher-order derivatives, which fundamentally prevents accurate evaluation of the required residuals and explains its poor performance. The tanh activation function, in contrast, is smooth and infinitely differentiable, satisfying the derivative requirements. However, in practice, as already mentioned, higher-order derivatives can lead to numerical instabilities, oscillations, or gradient scaling issues during training, which adversely affect convergence. For the $\tanh$ function in particular, Fig.~\ref{fig_activ_deriv} shows that the magnitude of the 6$^\mathrm{th}$-order derivative sharply increases near $x = 0$. Similar behavior is observed for the 4$^\mathrm{th}$ and 5$^\mathrm{th}$-order derivatives, which can induce strong variations in magnitude and, consequently, numerical instability. This can be also noted in \ref{sec_yoo} and Fig.~\ref{Fig_Yoo_tanh}. In contrast, the sine function derivatives merely alternates between sine and cosine forms, limiting its representation capability. Therefore, the choice of activation function must balance the existence of high-order derivatives with stable and efficient training behavior. Finally, the softplus and possibly sigmoid functions approximately capture the correct shape of the eigenmode, but they fail to satisfy the boundary conditions within the given number of epochs.

\begin{figure}
	\centering
	\includegraphics[width=10cm]{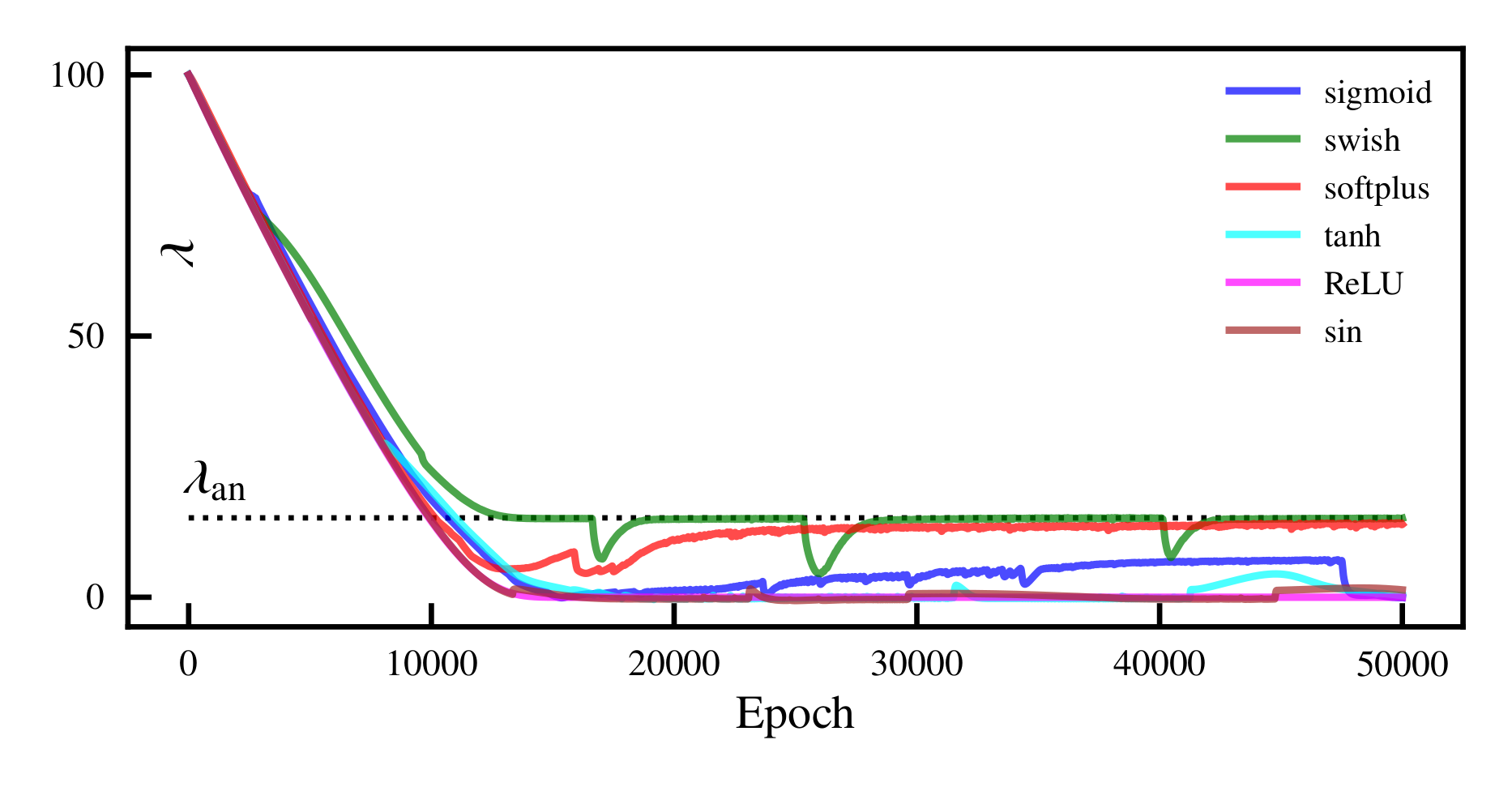}
	\caption{Effect of the choice of activation function on $\lambda$}
	\label{lambda_act_fun}
\end{figure}

\begin{figure}
	\centering
	\includegraphics[width=10cm]{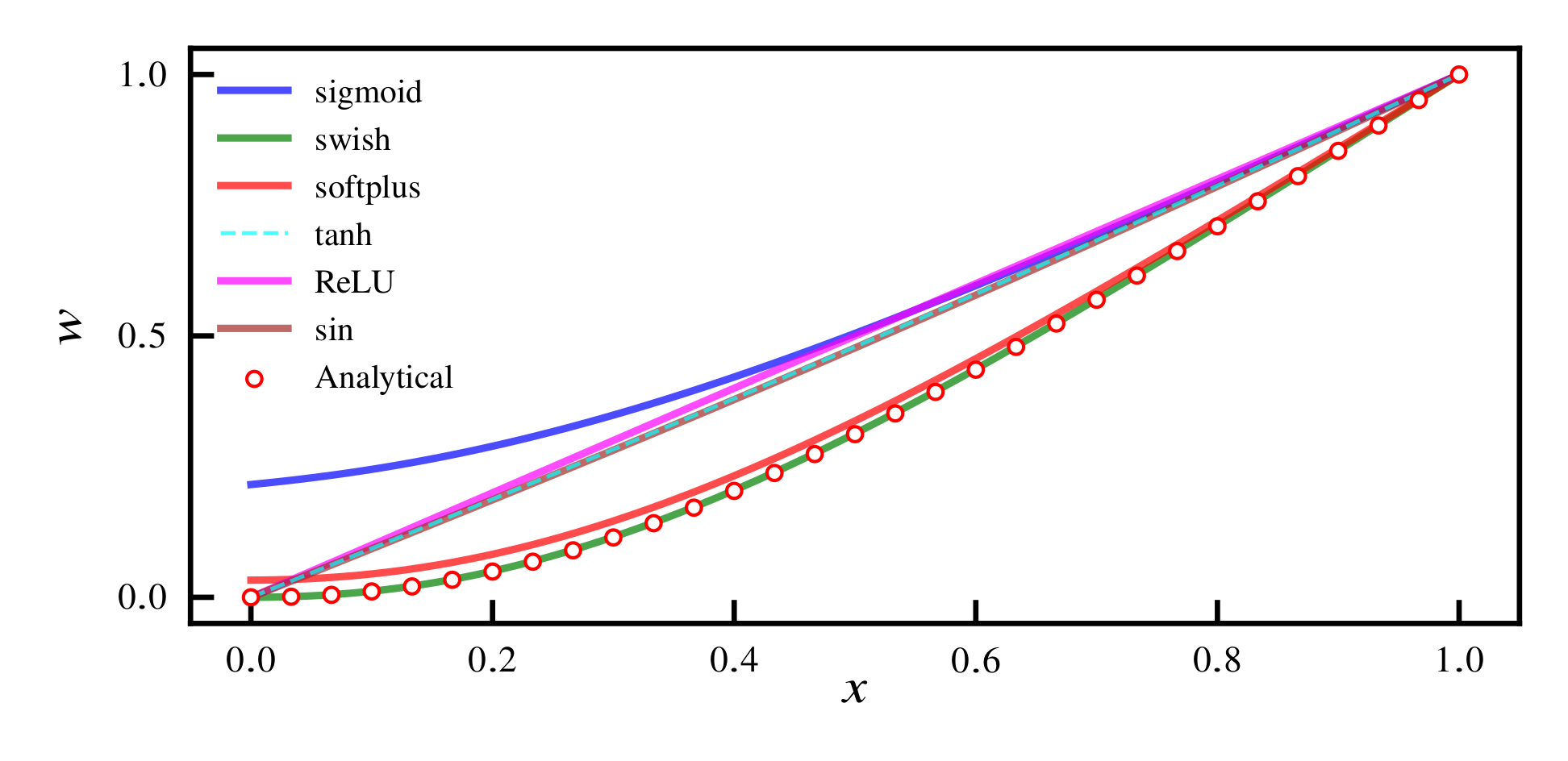}
	\caption{Effect of the choice of activation function on the first eigenmode}
	\label{mode_act_fun}
\end{figure}

\subsection{Results and Discussion}
\label{sec_case2_RandD}
Based on the above results, no gradient clipping was applied to any trainable variables in the final analysis, while an L2 regularization penalty of $10^{-6}$ was used. The results are presented in Tab.~\ref{tab_Nonlocal_CF} and Fig.~\ref{fig_CF_HR} for a high constant learning rate of $10^{-2}$. This higher learning rate enables more robust training; even when the initial guess for the eigenvalue ($\lambda_{\mathrm{ini}} = 100$) is far from the analytical value ($\lambda_{\mathrm{an}} = 15.1953$), the correct eigenvalue is obtained. The computational cost in this case is about 70 epochs per minute, which is 4.6 times slower than in the local case. Evidently, the calculation of sixth-order derivatives is significantly more involved than that of fourth-order derivatives. To illustrate training results for a lower and adaptive learning rate, an additional training is performed and results are documented in \ref{sec:app1}. The same computational cost is observed for the lower learning rate, and almost the same for the increased number of collocation points in \ref{sec:app3_higher}. 

Generally, during PINN training, higher learning rates are often used at the start to achieve faster convergence. Once the optimizer finds an initial solution, the learning rate is typically reduced to improve accuracy. In this particular case, during numerical experimentation, the higher learning rate showed less sensitivity to suboptimal choices of parameters and procedures (e.g., loss weights, L2 regularization, gradient clipping) compared to low and adaptive learning rates. This was primarily reflected in the failure of the lower learning rates to converge to the correct $\lambda$, although obtaining the first eigenfunction was less problematic. Training results for the low and adaptive learning rates are documented in \ref{sec:app1}, particularly in Fig.~\ref{fig_CF_LR}. It is evident that lower learning rates yield more accurate results, albeit at significantly increased computational cost. 

The initial values of $\lambda_{\mathrm{ini}}$ in ~\ref{sec:app1} were deliberately chosen to be close to the analytically obtained ones in order to reduce computational costs. The same strategy was used in \cite{Yoo2025}. However, it is not difficult to envisage a two-stage training scheme in which a higher learning rate is used at the beginning, and after a predefined period of patience - during which no improvement in the losses is observed, or some other criterion, the optimizer switches to a lower learning rate. In this way, the results presented in this section and in \ref{sec:app1} could be effectively combined in a single training. This approach would ensure generality in cases where the analytical value is not known a priori, while preserving the accuracy of the results.

Further, it appears that oscillations in the loss functions cannot be fully avoided with the presently available techniques. For example, see Fig. 3a in \cite{Yoo2025}, where a local beam eigenvalue problem governed by a fourth-order differential equation exhibits similar behavior. This phenomenon arises from the forward - inverse nature of the problem. However, when following the lower envelope of the loss curves (most clearly visible in Figs.~\ref{fig_CF_LR} and \ref{fig_SS_LR}, a clear convergence trend can be observed. So, with lower learning rates, the normalization losses also oscillate, but their magnitudes remain very small and thus not significant, indicating that the normalization constraint is enforced with high accuracy. For higher learning rates considered in this section, a similar pattern is observed, though with larger normalization losses and occasional abrupt jumps, after which the optimizer again tries to resume a downward trend.

\begin{table}
	\begin{center}
	\begin{tabular}{|l|r|}
		\hline
            No. epochs			  & \numprint{50000}    \\
		\hline
		Initial $\lambda$     & 100.0000   \\
		Predicted $\lambda$   & 15.1236  \\
		Analytical $\lambda$  & 15.1953	 \\
		Difference (\%) 	  &  0.4741  \\
		\hline
		min. $L_\mathrm{tot}$ & 0.5357 \\
		\hline
	\end{tabular}
	\caption{Number of training epochs, initial, predicted, analytical value of $\lambda$, and difference, total loss for a high learning rate for a nonlocal cantilever beam.}
	\label{tab_Nonlocal_CF}
	\end{center}
\end{table}

\begin{figure}
	\centering
	\includegraphics[width=11cm]{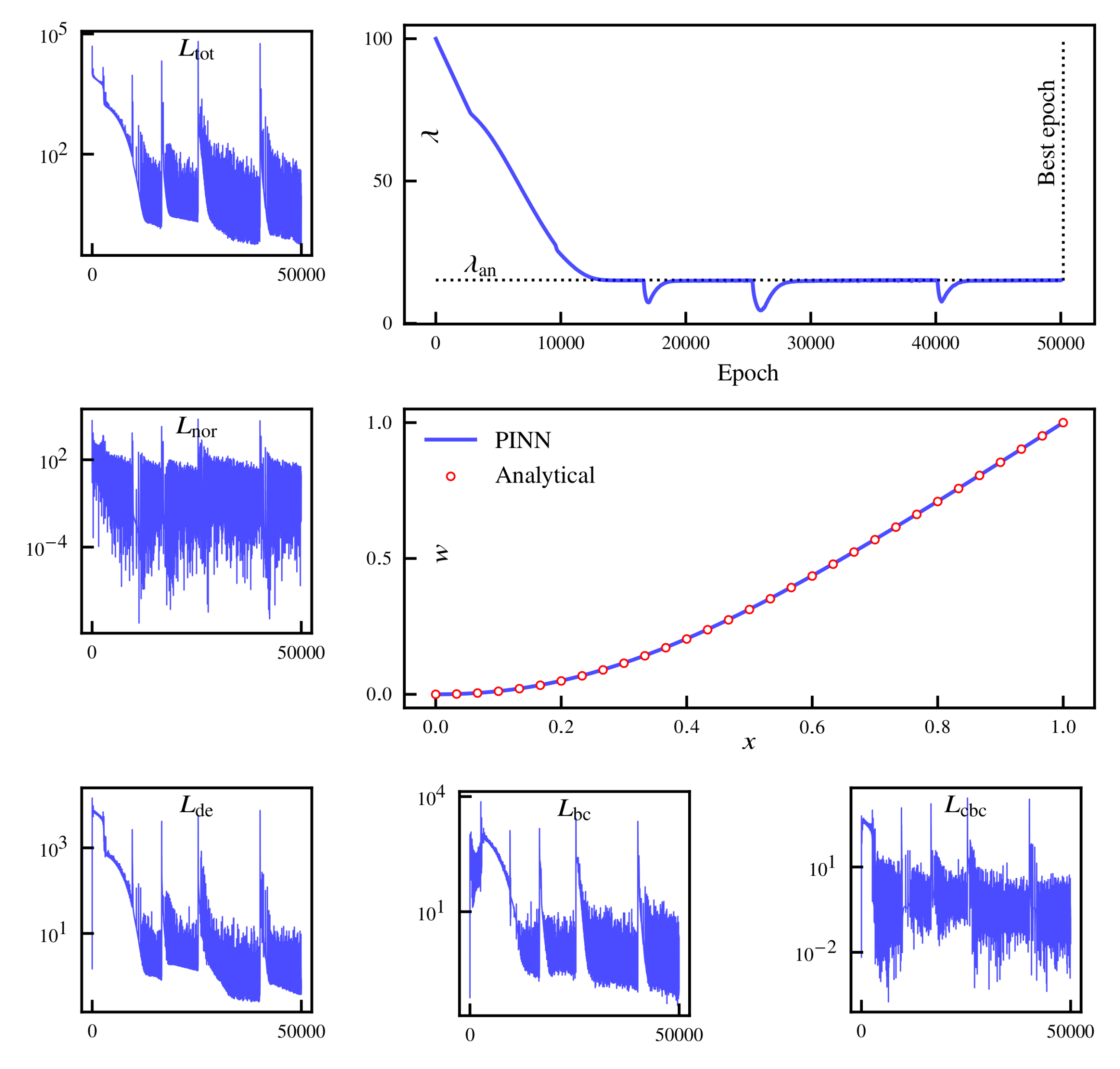}
	\caption{Nonlocal cantilever beam with a high learning rate: $\lambda$ convergence, first eigenfunction, and loss histories.}
	\label{fig_CF_HR}
\end{figure}

\subsection{Remarks on Convergence}
The convergence issue in PINNs is an important topic. In classical numerical methods, convergence can often be investigated through rigorous proofs. However, in stochastic optimization, which is used in PINNs, this is a highly complicated matter. In fact, in stochastic optimization \cite{Canadija2024}, the same simulation is typically run multiple times, and the best solution is selected at the end. The number of repetitions can start at 10, but 1000 repetitions are not uncommon. For this reason, in PINNs that employ stochastic optimization and automatic differentiation, it is very difficult to provide rigorous convergence proofs.

The convergence in the present case can be assessed by observing the following:
\begin{itemize}
\item Decrease of the total loss as well as other losses (de, bc, cbc, and nor) with the number of epochs.
\item Asymptotic behavior of $\lambda$ with respect to the known (analytical) values. In this respect, $\lambda$ can be determined in two ways: either by finding the minimum total loss when $\lambda$ has not yet reached an asymptotic value, or once $\lambda$ attains an asymptotic value.
\item Conformance (and stability) of the analytical eigenmode and the eigenmode predicted by the PINN.
\end{itemize}
If the above criteria are met, one can safely assume that the solution has converged to a physically consistent solution.

Unfortunately, due to the nonconvex nature of PINNs, this is not always the case, and convergence can occur to an incorrect or unwanted solution, frequently to a trivial local minimum. Factors influencing such outcomes include the choice of the initial point, initial weights and biases, as well as the choice of activation function, learning rate, and weighting factors, among others. To this end, two additional analyses were performed and presented in \ref{sec:app5_batches} and \ref{sec:app6_initial}.  In particular, it is shown that:
\begin{itemize}
\item Increasing the number of collocation points leads to more accurate solutions (see Appendix G).
\item In most cases, the choice of the initial value of $\lambda$ does not affect convergence to the correct solution (see Appendix H). In some instances where convergence is not achieved, starting from a different random weight initialization can resolve the issue in repeated simulations.
\end{itemize}
Consequently, although the results do not provide a definitive proof of convergence, nor guarantee that the method will converge in 100\% of simulations, they support the conclusion that the proposed method is sufficiently robust for the intended purpose.

\section{Case Study 3: PINN for Nonlocal Simply Supported Beam Eigenvalue Problem}
The final case considers the determination of the eigenstate for a nonlocal simply supported beam, as shown in Fig.~\ref{ssbeam}. All parameters and settings remain the same as for the nonlocal cantilever beam described in Sec.~\ref{CS_nonlocal_canti}, except for the boundary condition loss, which now takes the following form:
\begin{align}
	L_{\text{bc}} &= {w}(0)^2 + \left( \partial_{{x}}^2 {w}(0) - a^2 \partial_{{x}}^4 {w}(0) \right)^2 \notag \\
	&+  {w}(1)^2 + \left( \partial_{{x}}^2 {w}(1) - a^2 \partial_{{x}}^4 {w}(1) \right)^2.
	\label{kinematicbcSSloss}
\end{align}
The number of collocation points in this case is taken to be $N_\mathrm{c} = 101$ in order to include a point at the midspan of the beam, which is used for normalization of the eigenfunction.

% \begin{figure} 
% 		\centering
% 		\includegraphics[width=9cm]{nonlocal ss beam.png}
% 		\caption{Schematic of Nonlocal Simply Supported Beam}
% 		\label{ssbeam}
% \end{figure}
\begin{figure}
    \centering
    \includegraphics[width=0.7\linewidth]{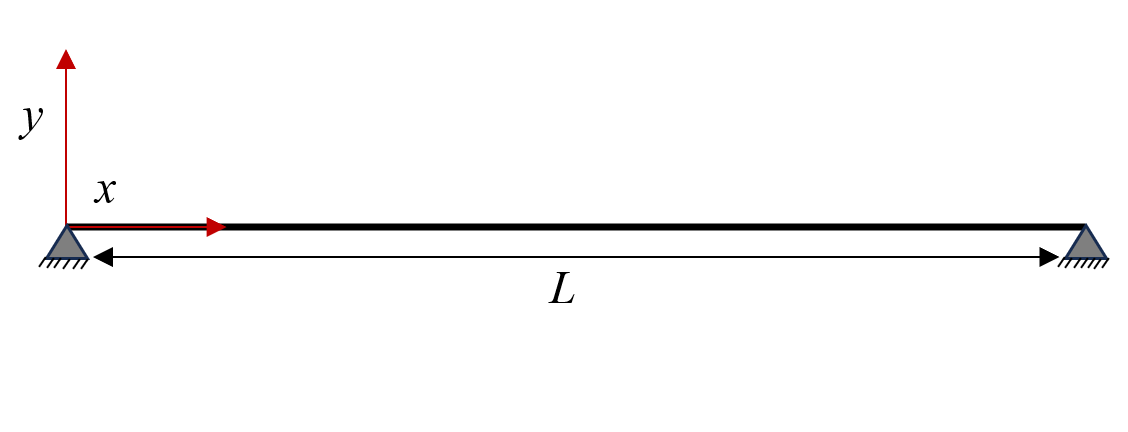}
    \caption{Schematic of Nonlocal Simply Supported Beam}
    \label{ssbeam}
\end{figure}

\subsection{Results and Discussion}
For the present case, the results are summarized in Tab.~\ref{tab_Nonlocal_SS} and Fig.~\ref{fig_SS_HR}. The same conclusions drawn for the nonlocal cantilever beam apply to the nonlocal simply supported beam. Once again, the training is robust and converges to the analytically obtained eigenstate. The problem is revisited using lower and adaptive learning rates in \ref{sec:app1}. During the trial-and-error process, it was again observed that the lower and adaptive learning rates are more sensitive to the choice of hyperparameters.

\begin{table}
	\begin{center}
	\begin{tabular}{|l|r|}
		\hline
            No. epochs			  & \numprint{80000}   \\
		\hline
		Initial $\lambda$     & 150.000	    \\
		Predicted $\lambda$   & 104.496  \\
		Analytical $\lambda$  & 105.133  \\
		Difference (\%) 	  &  0.6059 \\
		\hline
		min. $L_\mathrm{tot}$ & 0.1042 \\
		\hline
	\end{tabular}
	\caption{Number of training epochs, initial, predicted, analytical value of $\lambda$, and difference, total loss for a high learning rate for the nonlocal simply supported beam.}
	\label{tab_Nonlocal_SS}
	\end{center}
\end{table}

\begin{figure}
	\centering
	\includegraphics[width=11cm]{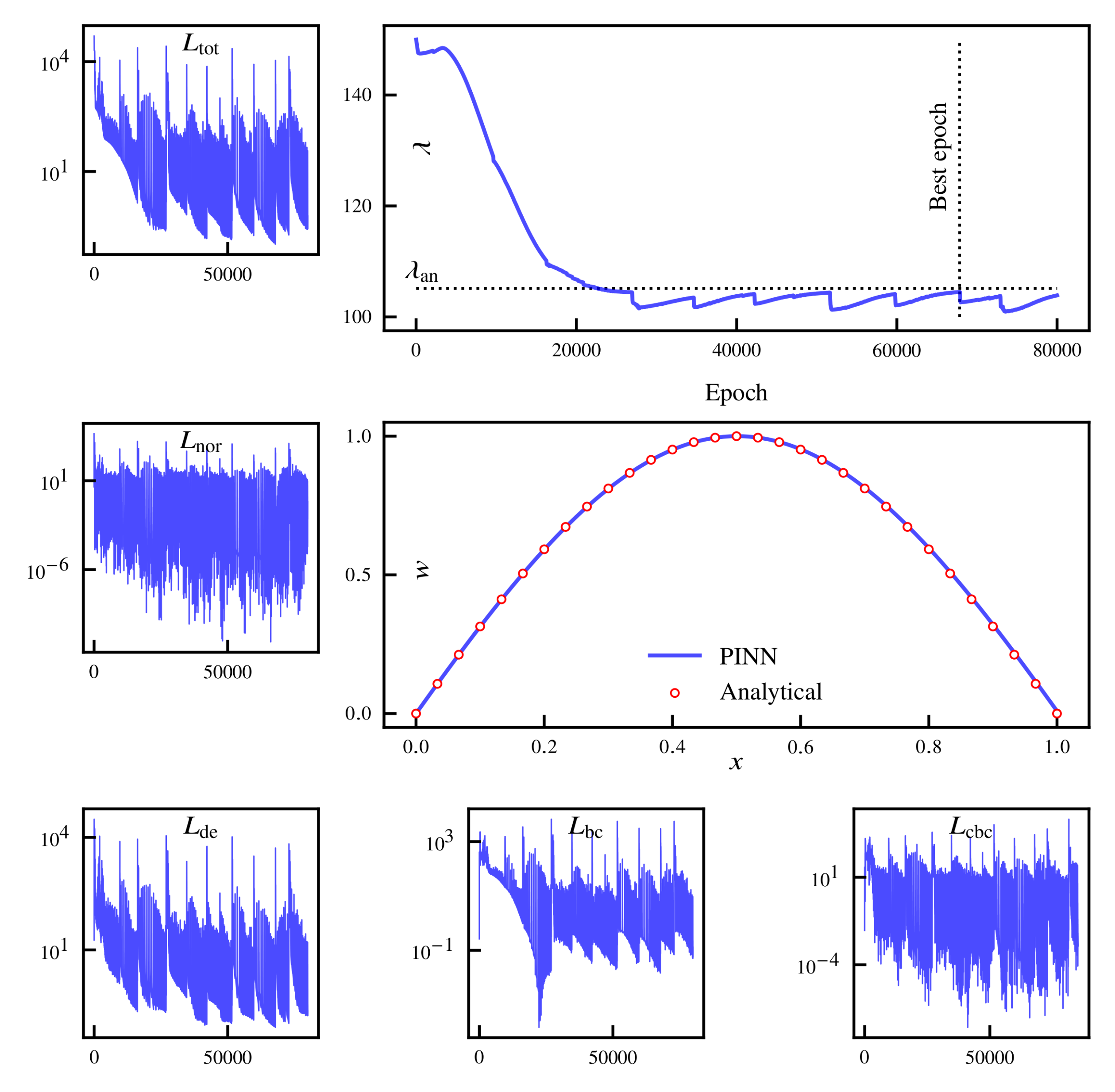}
	\caption{Nonlocal simply supported beam with a high learning rate: $\lambda$ convergence, first eigenfunction, and loss histories.}
	\label{fig_SS_HR}
\end{figure}

\section{Conclusions}
For the first time, this study presents a robust application of PINNs to address the eigenvalue problem associated with the free vibration of nonlocal cantilever and simply supported beams, as well as a local cantilever beam, for the determination of their first eigenvalues. As a preliminary step, the local problem was solved to efficiently obtain suitable hyperparameters, which were then reused for the nonlocal problem. By leveraging the stress-driven nonlocal model, the governing sixth-order differential equation was formulated and the unsupervised forward - inverse problem was solved using a PINN architecture that integrates the governing physical laws, along with kinematic and constitutive boundary conditions, directly into the loss function. The proposed methodology demonstrates high accuracy in predicting both the eigenvalues and corresponding eigenfunctions, achieving excellent agreement with analytical solutions across all considered structural configurations. 

It is also shown that a direct application of local eigenvalue solution procedures to the nonlocal problem is not straightforward. In particular, the commonly used tanh activation function was found to be unsuitable for sixth-order problems of this type. Moreover, enforcing the normalization condition over the entire domain, rather than at a single point, proved to be a more robust and accurate approach. Adaptive weighting procedures, such as the Relative Loss Balancing with Random Lookback Algorithm, which have demonstrated good performance in purely forward or inverse problems, did not yield satisfactory results in the present case. Regarding the reduced number of collocation points by batching and random selection of collocation points, it was found that this provides a slight increase in computational speed compared to the full-batch case, albeit at the expense of accuracy. Finally, the good convergence properties are confirmed by testing different initial values of $\lambda$ and by the improved accuracy of results with a larger number of collocation points.

The PINN approach offers significant advantages over traditional analytical and numerical methods, particularly in handling complex differential equations without the need for mesh generation or complex analytical procedures, which are also rather computationally intensive in the present case. The use of the swish activation function, carefully tuned hyperparameters, and a composite loss function incorporating differential equation residuals, boundary conditions, constitutive boundary conditions, and normalization terms ensured stable convergence and precise predictions. 

On the downside:
\begin{itemize}
	\item The determination of sixth-order derivatives via automatic differentiation is relatively slow and significantly prolongs the training process. However, once trained - as is typically the case with neural networks - the evaluation of the eigenstate is very fast.
	\item A change in boundary conditions, for example to a doubly clamped beam, requires retraining of the network. The same applies to a change in the underlying constitutive model, such as the two-phase stress-driven nonlocal theory. However, the nondimensional formulation used in the present case allows for simple recalculation during the postprocessing stage to account for different material or geometrical properties.
	\item A preliminary investigation into the determination of higher eigenstates shows that, although the present methodology can be applied, its computational efficiency is low and needs improvement. 
\end{itemize}

\section*{Data availability}
An example of code used in this research can be found at Das, Baidehi; Barretta , Raffaele; \v{C}ana\dj{}ija, Marko (2025), “PINN for Nonlocal Beam Eigenvalue Problems”, Mendeley Data, V1, doi: 10.17632/7hf2v6r237.1, \url{https://data.mendeley.com/datasets/7hf2v6r237/1}.
%\textit{Above data is still not published online. For the review purposes the preview can be found at: https://data.mendeley.com/preview/7hf2v6r237}  	
% https://data.mendeley.com/drafts/7hf2v6r237

\section*{Acknowledgments}
This work was supported by the project uniri-mzi-25-16 and Financial support from the Italian Ministry of University and Research (MUR) in the framework of the Project PRIN 2022 PNRR code P20223PLC2 "Nonlocal modelling of nano-coatings" is gratefully acknowledged.
%%%%%%%%%%%%%%%%%%%%%%%%%%%%%%%%
\begin{figure}[h]
    \centering
    \begin{minipage}{0.15\textwidth}
        \centering
        \includegraphics[width=\linewidth]{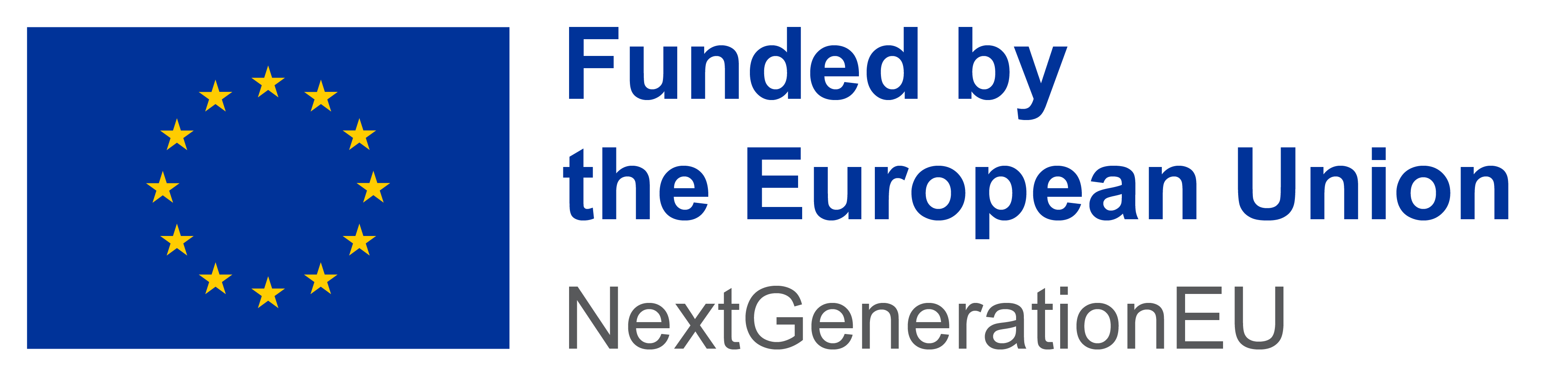}
        \label{fig:subA}
    \end{minipage}
    %\hfill
    \hspace{3pt}
    \begin{minipage}{0.1\textwidth}
        \centering
        \includegraphics[width=\linewidth]{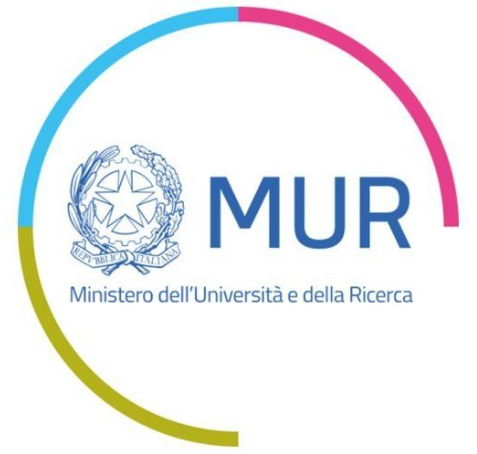}
        \label{fig:subB}
    \end{minipage}
    \label{fig:main}
\end{figure}

\appendix
\section{Results Obtained Using Lower and Adaptive Learning Rate}
\label{sec:app1}
In the following cases, exponential decay learning rates are used: one for neural network weights (initial rate $5 \times 10^{-4}$, decay steps \numprint{5000}, decay rate 0.9) and one for $\lambda$ (initial rate $1 \times 10^{-4}$, with the same decay parameters). For the nonlocal problems, the decay steps parameter was increased to \numprint{25000}.

The results are summarized in Tab.~\ref{tab_lowrate} and Figs.~\ref{fig_local_low_LR}, \ref{fig_CF_LR}, and \ref{fig_SS_LR} for the local cantilever beam, nonlocal cantilever beam, and nonlocal simply supported beam, respectively. For comparison, results for the simply supported beams at \numprint{150000} and \numprint{300000} epochs are also reported in Tab.~\ref{tab_lowrate}. All other hyperparameters were kept identical to those used in the high learning rate case. It is evident that the local cantilever beam requires the fewest epochs to converge. For the nonlocal beams, the cantilever beam achieves a lower total loss after \numprint{150000} epochs compared to the simply supported beam, although the difference in the predicted eigenvalues is larger for a cantilever beam. Both nonlocal beams exhibit relatively low loss at the end of training, and $\lambda$ can be obtained even closer to the analytical value if training is continued for additional epochs.
 
	\begin{table}
	\begin{center}
		\begin{tabular}{|l|r|r|r|r|}
			\hline
			& L CF  & NL CF & NL SS & NL SS \\
			\hline
			No. epochs			  & \numprint{50000} & \numprint{150000} & \numprint{150000} & \numprint{300000}\\
			\hline
			
			Initial $\lambda$     & 14 & 16 & 106 & 106 \\
			Predicted $\lambda$   & 12.3576 & 15.1813 & 105.088 & 105.111 \\
			Analytical $\lambda$  & 12.3596 & 15.1953 & 105.133 & 105.133\\
			Difference (\%) 	  &  0.016  &   0.092 &  0.043  &   0.021  \\
			\hline
			min. $L_\mathrm{tot}$ & 0.005068 & 0.07833 &  0.12374 & 0.06292 \\
			\hline
		\end{tabular}
		\caption{Number of training epochs, initial, predicted, and analytical values of $\lambda$, difference, and total loss for a low and adaptive learning rate for a local cantilever beam (L CF), nonlocal cantilever beam (NL CF), and nonlocal simply supported beam (NL SS). For the simply supported beam, results for the same training at \numprint{150000} and \numprint{300000} epochs are reported.  }
		\label{tab_lowrate}
		\end{center}
	\end{table}
	\begin{figure}
		\centering
		\includegraphics[width=11cm]{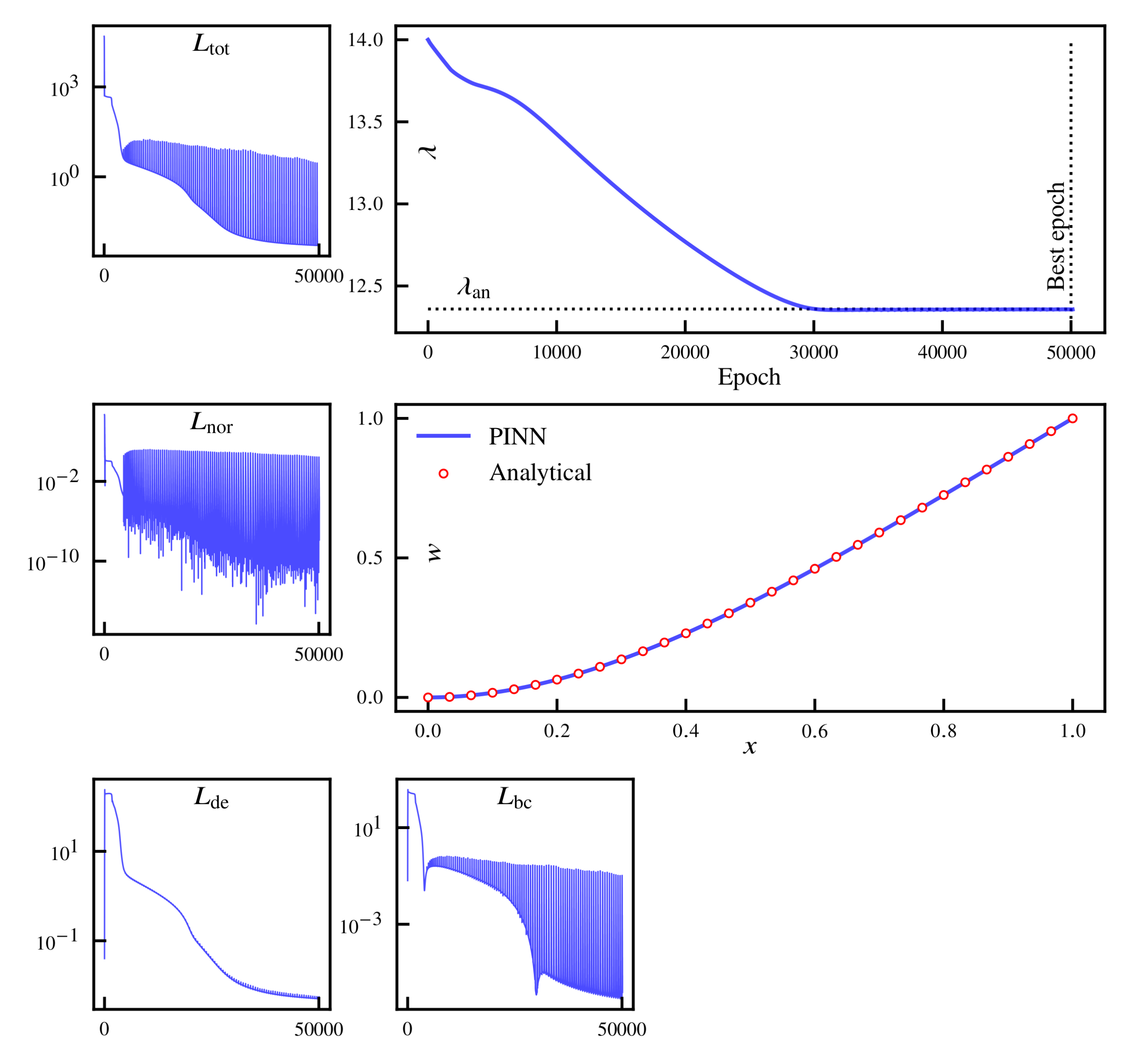}
		\caption{Local cantilever beam with low and adaptive learning rate: $\lambda$ convergence, first eigenfunction and loss histories.}
		\label{fig_local_low_LR}
	\end{figure}
	
	\begin{figure}
		\centering
		\includegraphics[width=11cm]{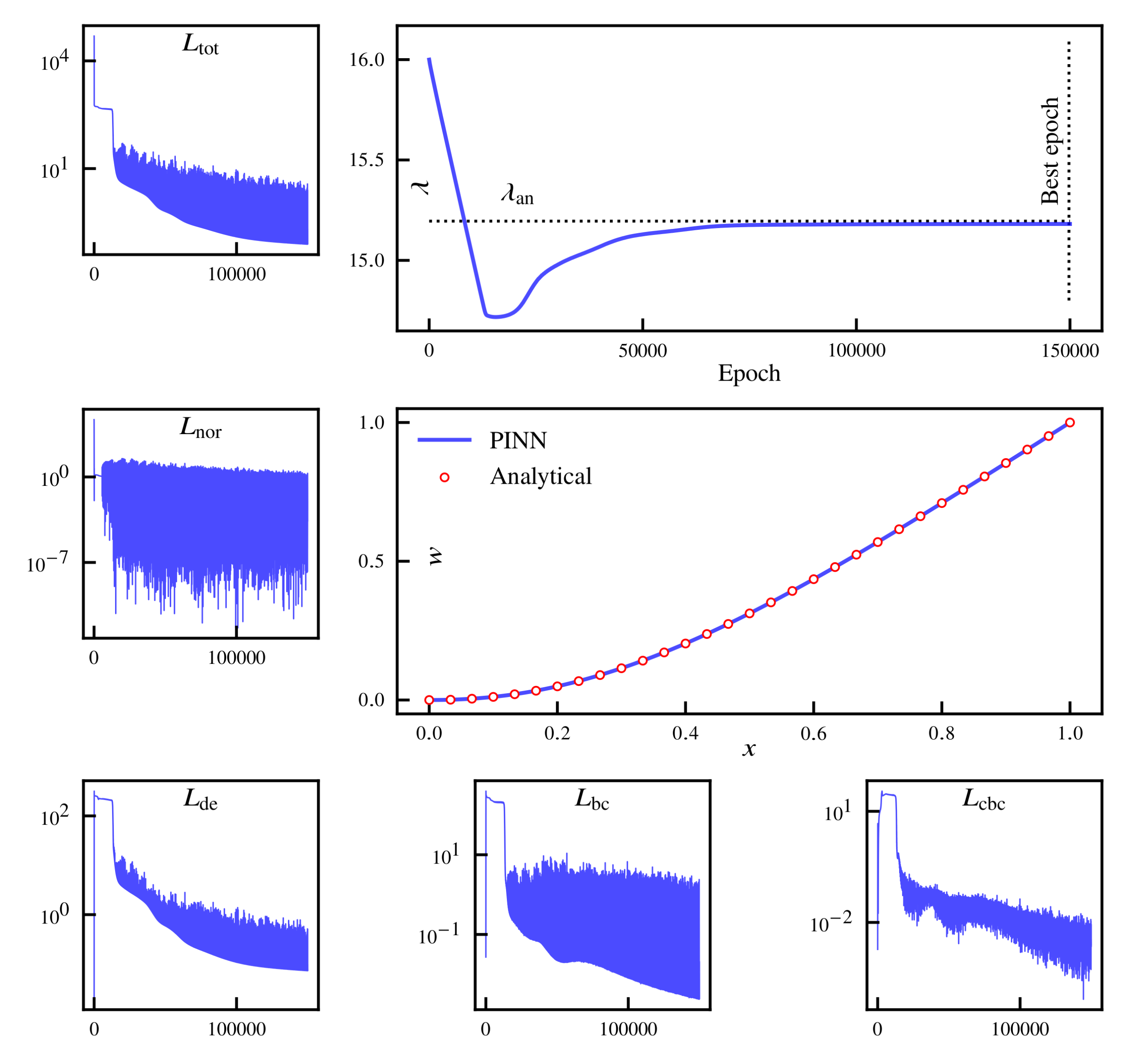}
		\caption{Nonlocal cantilever beam with low and adaptive learning rate: $\lambda$ convergence, first eigenfunction and loss histories.}
		\label{fig_CF_LR}
	\end{figure}

	\begin{figure}
	\centering
	\includegraphics[width=11cm]{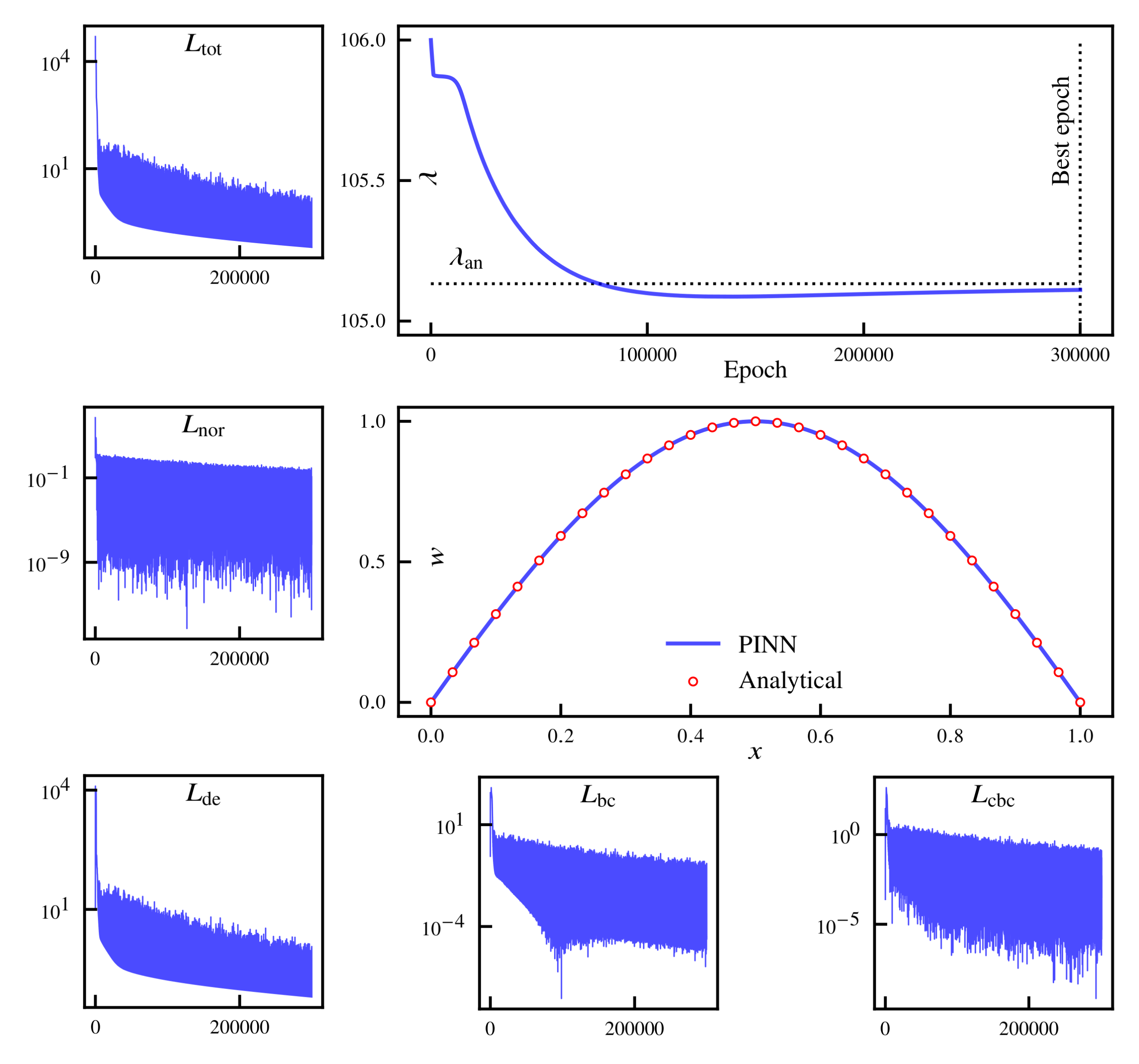}
	\caption{Nonlocal simply supported beam with low and adaptive learning rate: $\lambda$ convergence, first eigenfunction and loss histories.}
	\label{fig_SS_LR}
\end{figure}
\newpage

\section{Analytical Solution of the Nonlocal Beam Eigenvalue Problem}

\label{sec:app2}

The non-dimensional governing equation  as provided in Eq.~\eqref{nondim} is a sixth order differential equation equipped with four classical and two constitutive boundary conditions.
To solve this analytically for free vibration, a harmonic solution in time is assumed:

\begin{equation}
   \bar{w}(\bar{x}, t) = \phi(\bar{x}) \cos(\omega t). 
   \label{harmonicsol}
\end{equation}

This leads to $\partial^2 \bar{w} / \partial t^2 = -\omega^2 \phi(\bar{x}) \cos(\omega t)$. Substituting and dropping the time part for the amplitude equation yields the spatial solution as:

\begin{equation}
  a^2 \frac{d^6 \phi}{d \bar{x}^6} - \frac{d^4 \phi}{d \bar{x}^4} - \lambda \phi = 0.
  \label{analyticaleigeq}
\end{equation}

Assuming a solution of the form $\phi(\bar{x}) = e^{r \bar{x}}$  and substituting into Eq.~\eqref{analyticaleigeq}  gives the characteristic equation:
\begin{equation}
a^2 r^6 - r^4 - \lambda = 0.
\label{characeq}
\end{equation}

Let $s = r^2$, where $r$ denotes the roots of the characteristic equation. Substituting Eq.~\eqref{characeq} using this relation could be transformed into a cubic polynomial:

\begin{equation}
a^2 s^3 - s^2 - \lambda = 0.
\label{characeq2}
\end{equation}
\indent To solve this cubic equation analytically, Cardano's formula has been used. This involves a primary step to depress the cubic polynomial by eliminating the quadratic part by the effective substitution of $s = z + \dfrac{1}{3a^2}$ and the depressed form is given as:
\begin{equation}
a^2 z^3 - \dfrac{z}{3a^2} - \dfrac{2}{27a^4} - \lambda = 0\,,
\label{depressedform1}
\end{equation}
which could be simply written as:
\begin{equation}
a^2z^3 + P z + Q = 0,
\label{depressedform2}
\end{equation}
where
$P = -\dfrac{1}{3a^2}$,
$Q = -\dfrac{2}{27a^4} - \lambda\,.$

\indent Let us consider the position $z = u + v$ where $u$ and $v$ must satisfy the following conditions
\begin{equation}
    \begin{cases}
   u^3 + v^3  = - Q \\
   u\,v = -\displaystyle\frac{P}{3}\,
    \end{cases}
    \label{uvroots}
\end{equation}
The general solution is expressed as
$$u = \sqrt[3]{-\frac{Q}{2} + \sqrt{\Delta}}, \quad v = \sqrt[3]{-\frac{Q}{2} - \sqrt{\Delta}},$$
with $\Delta = \displaystyle\left(\frac{Q}{2}\right)^2 + \left(\frac{P}{3}\right)^3$ and $s = z + \dfrac{1}{3a^2}$.
For each $s_i$, compute $r = \pm \sqrt{s_i}$ (real roots) or $r = \pm i \sqrt{|s_i|}$ (imaginary roots). With six roots ($r_1$ to $r_6$), the general solution is:
\begin{equation}
\phi(\bar{x}) = \sum_{k=1}^6 c_k e^{r_k \bar{x}}
\label{generalsolana}
\end{equation}
with $\bar{x} \in [0, 1]$ and $c_k$ are constants.
Standard and non-standard (constitutive) boundary conditions must be prescribed. For instance, let us consider the case of nonlocal cantilever beam, whose boundary conditions are:
\begin{equation}
 \begin{cases}
\phi(0) = 0, \phi'(0) = 0, \phi'''(0) - \dfrac{1}{a} \phi''(0) = 0, \\
\phi''(1) - a^2 \phi''''(1) = 0, \phi'''(1) - a^2 \phi^{(5)}(1) = 0, \phi'''(1) + \dfrac{1}{a} \phi''(1) = 0\,.
\end{cases}
\end{equation}
\indent A homogeneous linear algebraic system is thus obtained, that can be also written in the following matrix form:
\begin{equation}
    \mathbf{M}(\lambda,\,a) \,\mathbf{c} = 0\,.
\end{equation}
where $c$ is a six-dimensional array of constants and $\mathbf{M}$ is the coefficient matrix. Natural frequencies are obtained by solving the following nonlinear equation in $\lambda$
\begin{equation}
    \det(\mathbf{M}_a(\lambda))=0
\end{equation}
for any given positive non-dimensional nonlocal parameter $a$. For further details and particular analytical solutions, the reader is refereed to \cite{apuzzo2017free}.

\section{Activation Functions and their Derivatives}
\label{App_activation}
Since the present calculations involve higher-order derivatives of activation functions, this section presents the activation functions used, along with their derivatives up to the 6$^\mathrm{th}$ order (Fig.~\ref{fig_activ_deriv}). It can be observed that the magnitudes of the derivatives of the $\tanh$ function increase significantly with the derivative order. Likewise, the ReLU activation function does not have derivatives of the 2$^\mathrm{nd}$ and higher orders.

	\begin{figure}
	\centering
	\includegraphics[width=12cm]{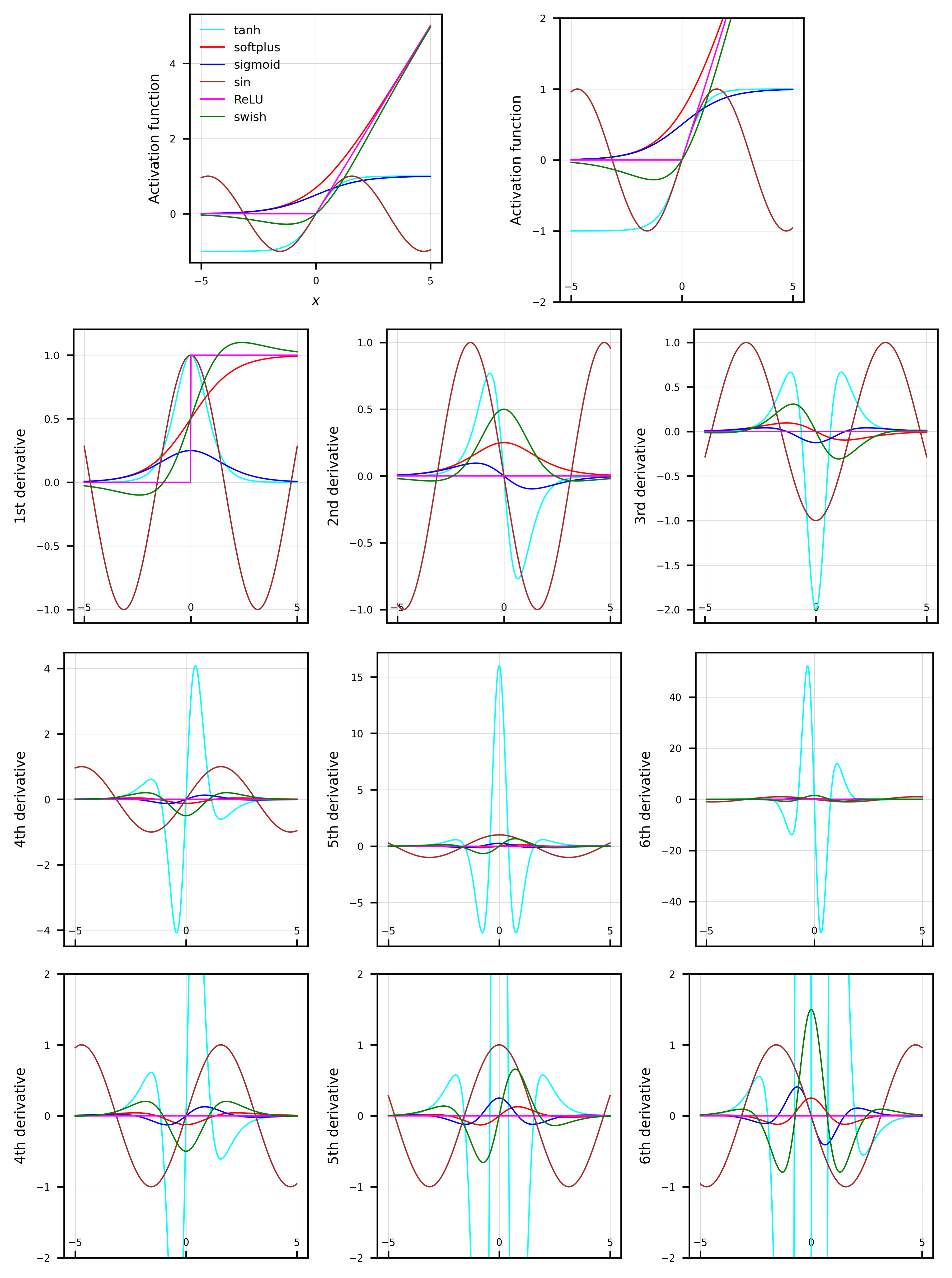}
	\caption{Activation functions (first row) and their derivatives up to the 6$^\mathrm{th}$ order. The bottom row shows zoomed-in views.}
	\label{fig_activ_deriv}
	\end{figure}

\section{Comparison with Yoo et al. \cite{Yoo2025}}
\label{sec_yoo}
The present section partially relies on the PINN model presented in \cite{Yoo2025}. The hyperparameters are taken to be the same as in Sec.~\ref{CS_nonlocal_canti}, but the integral normalization loss enforced over the entire domain, as defined in Eq.~(\ref{normloss}), is replaced by a pointwise normalization method. Specifically, the amplitude scaling in \cite{Yoo2025} for the local beam problem governed by a fourth-order differential equation is used:
\begin{equation}
	\label{normloss_point}
	L_\mathrm{nor}= \left| w(x_\mathrm{a}) - w_\mathrm{a} \right| ^2,
\end{equation}
where $w(x_\mathrm{a})$ is the vertical displacement predicted by the PINN at the point with abscissa $x_\mathrm{a}$ at which the transverse displacement $w_\mathrm{a}$ is prescribed. In the cantilever beam considered here, $x_\mathrm{a}=1$  and $w_\mathrm{a}=1$. Since this represents a prescribed displacement, it is logical to use the same weight factor as for the boundary conditions, i.e., $\alpha_{\mathrm{nor}}=100$. Training is performed using two activation functions: swish (Fig.~\ref{Fig_Yoo_swish}) and tanh (Fig.~\ref{Fig_Yoo_tanh}). The original method in \cite{Yoo2025} employed tanh as the activation function.

The results for the swish function in Fig.~\ref{Fig_Yoo_swish} indicate that approximately the same number of epochs is required to reach the correct $\lambda$ for both normalization strategies, i.e. integral normalization given in Eq.~(\ref{normloss}) and the pointwise normalization in Eq.~(\ref{normloss_point}). The convergence of $\lambda$ for the integral normalization appears slightly better than for the pointwise approach. However, when tanh activation function is considered, the situation is markedly different. Although tanh performs well in the case of the fourth-order problems \cite{Yoo2025}, the unfavorable properties of its sixth derivative, as demonstrated in \ref{App_activation} and Sec.~\ref{sec_CS2_preliminary} cause difficulties in the present sixth-order problem. As a result, neither the correct eigenmode nor the correct eigenvalue is obtained (Fig.~\ref{Fig_Yoo_tanh}). It is also worth noting the pronounced jumps and the large magnitudes of the losses.

\begin{figure}
	\centering
	\includegraphics[width=11cm]{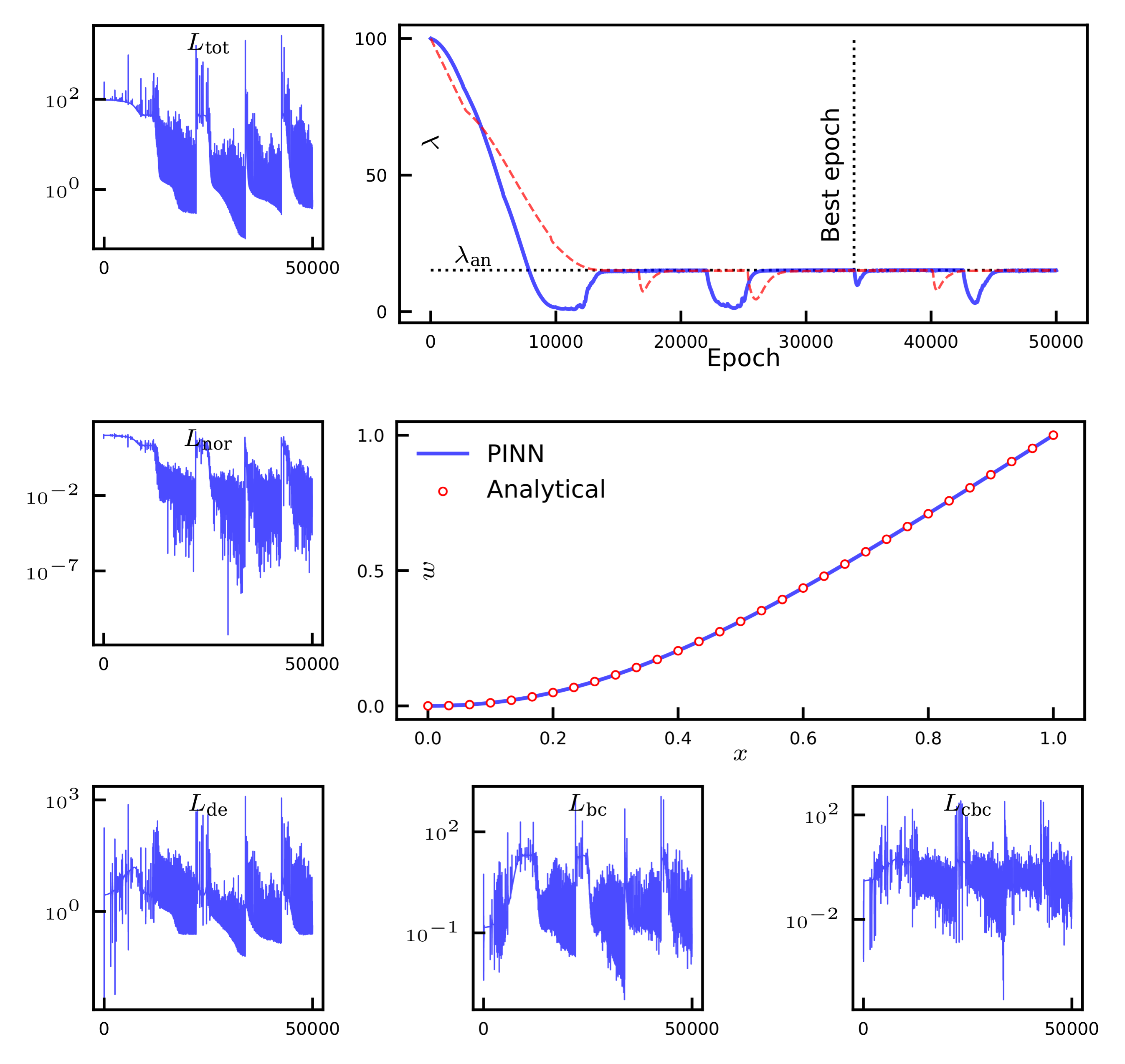}
	\caption{Results obtained with a point normalization (blue) using swish activation function: nonlocal cantilever beam with a high learning rate: $\lambda$ convergence, first eigenfunction, and loss histories. Red dashed line are results from Fig.~\ref{fig_CF_HR}. }
	\label{Fig_Yoo_swish}
\end{figure}

\begin{figure}
	\centering
	\includegraphics[width=11cm]{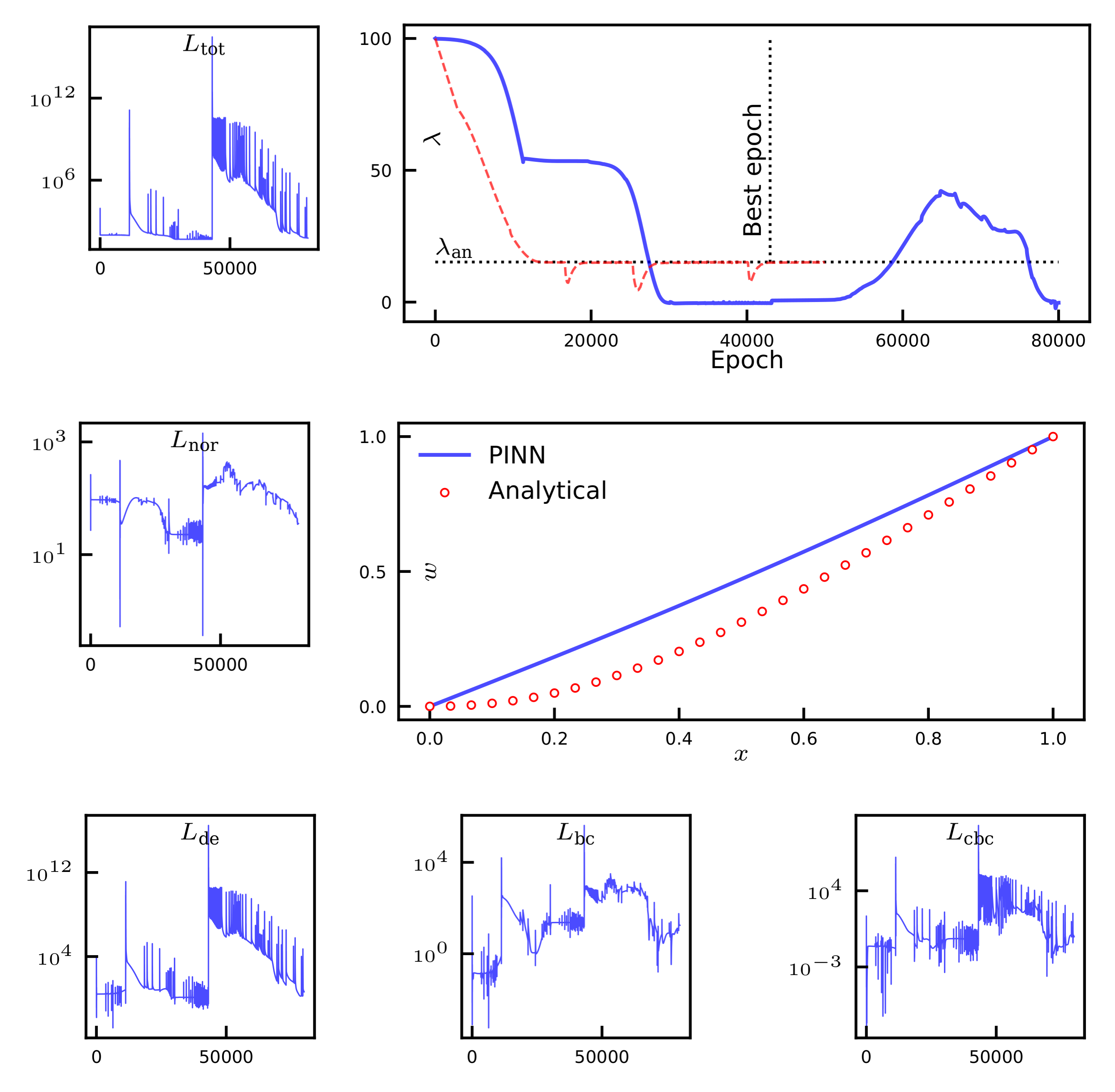}
	\caption{Results obtained with a point normalization (blue) using tanh activation function: nonlocal cantilever beam with a high learning rate: $\lambda$ convergence, first eigenfunction, and loss histories. Red dashed line are results from Fig.~\ref{fig_CF_HR}.}
	\label{Fig_Yoo_tanh}
\end{figure}

\section{Higher modes}
\label{sec:app3_higher}
The proposed methodology cannot be directly applied to determine higher modes, as there is an additional issue that need to be addressed.  Neural networks favor lower frequencies \cite{rahaman2019spectral}, so the natural tendency is to converge toward lower eigenmodes. In the case of the fourth-order problem \cite{Yoo2025}, this is addressed by introducing a new loss term that constrains an eigenvalue to be above a given threshold. A similar approach was pursued in \cite{jin2020unsupervised}. Consequently, both approaches rely on the use of a non-physical term. In \cite{Yoo2025}, the method still requires starting from a point close to the anticipated eigenvalue, using lower learning rates ($10^{-5}$), and a large number of epochs ($\numprint{400000}$ for the third mode) to converge to the correct eigenvalue. More physical approach \cite{Jin2022, HARCOMBE2023102136} uses the orthogonality condition, which must also be satisfied. To this end, an orthogonality loss is introduced as follows:
\begin{equation}
	L_{\text{orth}} = \sum_{i=1}^{M} \left( \int_{0}^{L} ( w_i) w  \, dx \right)^2 ,
	\label{loss_orth1}
\end{equation}
where $M$ is the total number of already computed eigenmodes $w_i$, and \( w _j\) is the predicted eigenfunction. The midpoint rule then provides the following approximation:
\begin{equation}
	L_{\mathrm{orth}} \approx  \sum_{i=1}^{M} \left(\sum_{j=1}^{N_\mathrm{c}-1} (w_i w_j) \right)^2,
	\label{orthloss_approx}
\end{equation}
This loss is multiplied by a weighting factor and added as $\alpha_{\mathrm{orth}} L_{\text{orth}}$ to the total loss, see Eq.~(\ref{totalloss}). Typically, identical weighting factors are used for both the orthogonality and normalization losses. Note that if the normalization loss from \cite{HARCOMBE2023102136} (see Eq.~(\ref{normloss_approx_Harcombe})) is used, one must retain $\Delta x$ in order to apply $\alpha_{\mathrm{orth}} = \alpha_{\mathrm{nor}}$, or least that these weights are of the similar order of magnitude.
\begin{equation}
	L_{\mathrm{orth}} \approx  \sum_{i=1}^{M} \left(\sum_{j=1}^{N_\mathrm{c}-1} (w_i w_j) \Delta x \right)^2.
	\label{orthloss_approx2}
\end{equation} 
Since this paper is primarily focused on the first mode, the present results should be regarded as preliminary and warrant further investigation to improve the quality of the solution. No effort has been made to optimize the weighting factor for the orthogonality loss term. In addition, no constraint loss term has been used to prevent convergence toward lower eigenmodes in the sense of \cite{jin2020unsupervised,Yoo2025}.

For the second mode, the same settings as in Section~\ref{sec:app1} were used, with an updated weighting factor of $\alpha_\text{nor} = 50$. Number of collocation points was 200. Although a large number of epochs is required, it is evident that the eigenmode computed by the PINN is already visually indistinguishable from the analytical solution at epoch \numprint{10000} (Fig.~\ref{Fig_2nd_mode}c). However, the calculation of the corresponding eigenvalue remains the most computationally demanding part. A more accurate eigenvalue could be obtained if the training were allowed to continue further. In other words, solving the forward part of the problem (the eigenmode) is relatively easy, whereas the inverse part (the eigenvalue) is significantly more challenging. The tabulated results in Tab.~\ref{tab_lowrate2ndmode} also demonstrate good accuracy. Although the number of collocation points has doubled relative to Sec.~\ref{sec_case2_RandD}, the computational cost remained almost unchanged at approximately 69 epochs per minute.

In the case of the third mode, $\alpha_\text{nor} = 500$, while the decay step of the learning rate was set to \numprint{75000} to account for the larger number of expected epochs. Similar to the second mode, the visually correct eigenmode is obtained rather early in the simulation, whereas the determination of $\lambda$ remains computationally expensive. It is clear that $\lambda$ still did not converged properly.

\begin{figure}
	\centering
	\includegraphics[width=14cm]{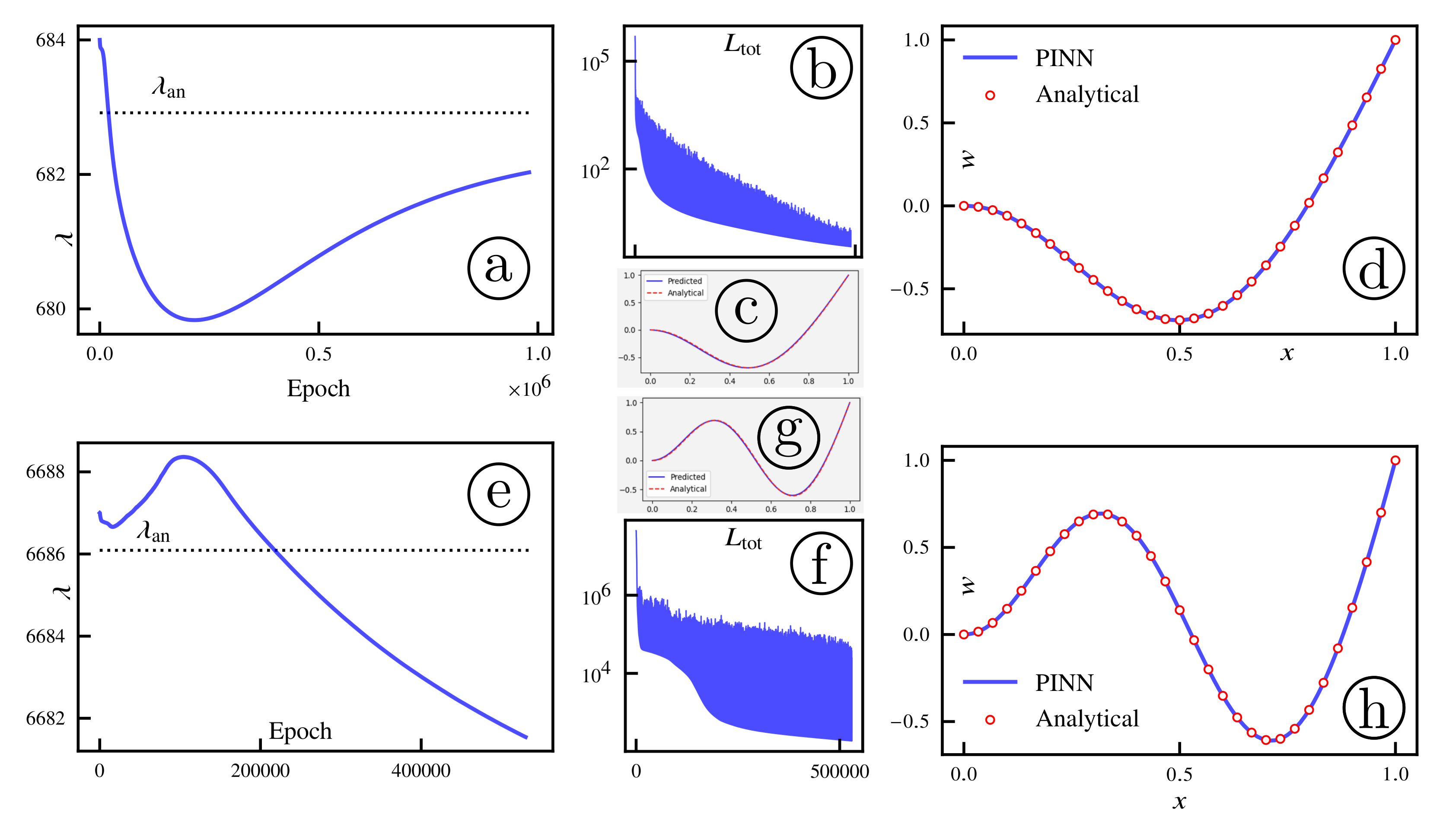}
	\caption{The second (a-d) and the third (e-h) eigenstates of the nonlocal cantilever beam with low and adaptive learning rates: $\lambda$ (a, e) convergence, eigenfunction (d, h), total loss (b, f) histories, and eigenmodes at epoch \numprint{10000} (c) and at epoch \numprint{14000} (g).}
	\label{Fig_2nd_mode}
\end{figure}

	\begin{table}
	\begin{center}
		\begin{tabular}{|r|r|r|}
			\hline
			& 2$^\mathrm{nd}$ mode & 3$^\mathrm{rd}$ mode  \\
			\hline
			No. epochs			  & \numprint{985000} & \numprint{537500}\\
			\hline
			
			Initial $\lambda$     & 684.000 & 6687.00\\
			Predicted $\lambda$   & 682.031 & 6681.54$^{*}$ \\
			Analytical $\lambda$  & 682.916 & 6686.09\\
			Difference (\%) 	  &  0.130  & 0.068$^{*}$\\
			\hline
			min. $L_\mathrm{tot}$ & 0.6904  & 187.08\\
			\hline
		\end{tabular}
		\caption{Number of training epochs, initial, predicted, and analytical values of $\lambda$, difference, and total loss for a low and adaptive learning rate for the second and the third eigenmode of the nonlocal cantilever beam. $^{*}$still not converged.}
		\label{tab_lowrate2ndmode}
	\end{center}
\end{table}

\section{Adaptive Selection of Weights}
\label{sec:app4_adaptive}
	Manual tuning of the weights in Eq.~(\ref{totalloss}) is a laborious and problem-dependent task. Recently, a procedure to automatically determine these weights was proposed in \cite{bischof2025multi}, known as the Relative Loss Balancing with Random Lookback Algorithm (ReLoBRaLo), and further refined for robustness in \cite{lourencco2025use}. In this approach, the weights are updated dynamically at each epoch based on a moving average of the losses combined with a random lookback mechanism. Thus, for the loss term $i$ (where $i$ represents de, bc, and nor weights in the local case, and additionally cbc in the nonlocal case) at epoch $j$, the weights are:
\begin{equation}
	\alpha_i^j=\xi \left( \rho \alpha_i^{j-1} + (1-\rho) \hat{\alpha}_i^{(j;0)} \right) + (1-\xi) \hat{\alpha}_i^{(j;j-1)}
	\label{eq_adapt1}
\end{equation}
where
\begin{equation}
	\hat{\alpha}_i^{(j;k)}=m \frac{\exp\left(\frac{L_i^j}{T \, L_i^k +\epsilon}-\max\frac{L_i^j}{L_i^k}\right)}{\sum_{p=1}^{m} \exp\left(\frac{L_p^j}{T \, L_p^k +\epsilon}-\max\frac{L_p^j}{L_p^k}\right) }.
	\label{eq_adapt2}
\end{equation}
	Here, $m$ is the total number of loss terms ($m=3$ in the local case and $m=4$ in the nonlocal case), $\epsilon = 10^{-8}$ is a small constant used to avoid division by zero, $\rho$ is a Bernoulli random variable (with $\mathbb{E}[\rho] \approx 1$) used in the random lookback, and $\xi$ is a parameter describing the influence of past values, usually chosen between 0.9 and 0.999. Setting $\mathbb{E}[\rho]=0$ results in always using the initial training loss, whereas $\mathbb{E}[\rho]=1$ uses only the most recent value of the loss term $i$. The hyperparameter $T$ is the so-called temperature, typically taken in the range $\left[10^{-6}, 10^2\right]$.

	The present case considers a simpler local cantilever beam. The above parameters were obtained through a simple grid search, resulting in $\xi = 0.999$, $T = 1$, and $\mathbb{E}[\rho] = 0.99$. The corresponding results are shown in Fig.~\ref{Fig_adaptive}. It is evident that the adaptive weight selection performs significantly worse than manual tuning of the weights - even the eigenmode is not captured correctly. While ReLoBRaLo has been known to perform well for purely forward or inverse problems, this is not the case for the forward–inverse problem considered here. The total number of epochs was \numprint{89000}, with the best epoch at \numprint{80940}, corresponding to a minimal loss of 0.03153 and a predicted $\lambda = 10.9803$ ($\lambda_\mathrm{an} = 12.3596$).

\begin{figure}
	\centering
	\includegraphics[width=11cm]{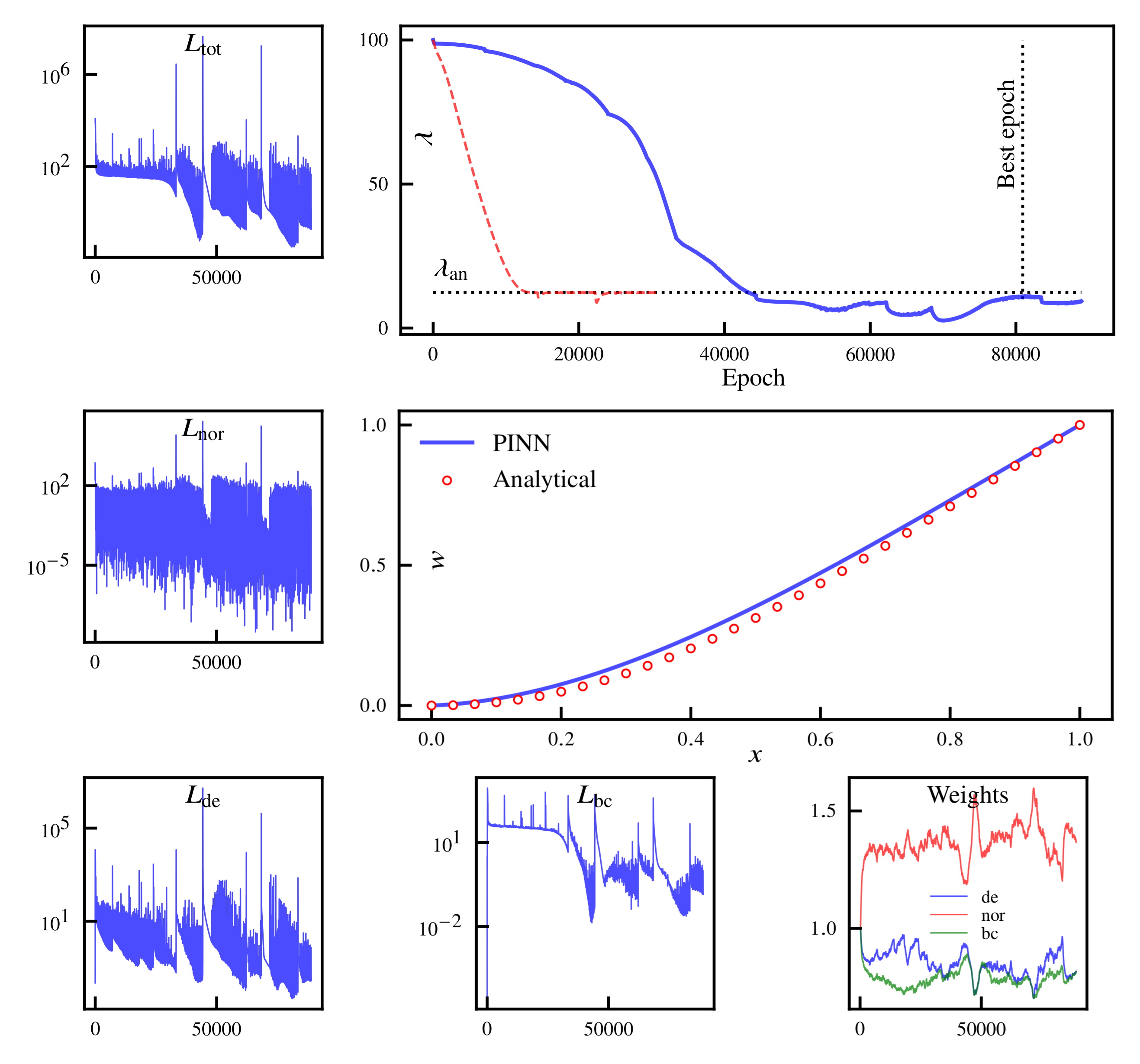}
	\caption{Local cantilever beam with learning rate $10^{-2}$: $\lambda$ convergence, first eigenfunction, loss and weights histories.}
	\label{Fig_adaptive}
\end{figure}

\section{Batching and Random Selection of Collocation Points}
\label{sec:app5_batches}
Since the evaluation of sixth-order derivatives is computationally very expensive, it is more efficient to compute these derivatives for the entire batch rather than invoke them multiple times. The latter approach would be prohibitively slow. For this reason, an alternative strategy is adopted: in each epoch, a specified number of randomly selected collocation points is used, and this selection is perturbed in every epoch. The only exception were two boundary points. The same procedure was followed in \cite{Jin2022}. Accordingly, out of the originally used 100 collocation points, batches of 10, 40, and 70 are employed.

However, the random selection of collocation points implies that the normalization loss defined by Eq.~(\ref{normloss_approx}) cannot be used, since $\Delta x$ is no longer constant as in the case of a uniform point distribution. For this reason, one must rely on:
\begin{equation}
	L_{\mathrm{nor}} \approx L_{\mathrm{nor}} \approx \left(\sum_{i=1}^{N_\mathrm{c}-1} (w_m)_i^2\Delta x_i - 1 \right)^2,
	\label{normloss_approx_Harcombe}
\end{equation}
used in \cite{HARCOMBE2023102136} and described in Sec.~\ref{sec_pinn_met}. Adopting this approach allows for different $\Delta x_i$ values at each midpoint. As discussed in Sec.~\ref{sec_pinn_met}, it also requires a modification of the normalization weight, which is now taken as $\alpha_{\mathrm{nor}} = \numprint{25000}$. The learning rate was $10^{-2}$. All other hyperparameters remain unchanged. For a fair comparison, the full batch of 100 uniformly distributed collocation points is also recalculated using the loss function given by Eq.~(\ref{normloss_approx_Harcombe}).

The obtained results are shown in Fig.~\ref{Fig_batches}. For comparison, earlier results obtained using a uniform distribution of 100 collocation points with the normalization loss Eq.~(\ref{normloss_approx}), as reported in Tab.~\ref{tab_Nonlocal_CF}, are also provided. Apart from the lowest number of collocation points ($N_\mathrm{c}=10$), all other choices resulted in the correct eigenvalue. Inspection of the numerical values in Tab.~\ref{tab_Nonlocal_CF_batch} reveals a clear trend of improvement in both the eigenvalues and the minimal losses as the number of collocation points increases, confirming the good convergence properties of the method toward both the correct eigenvalue and eigenmode. Finally, although at this point no attempt was made to distinguish between the influence of batching and random selection, it is known that in problems of this kind, random selection of points actually makes the solution worse \cite{kianian2025pinn}.

Finally, the computational cost for a batch size of 10 is 82.1 epochs per minute, which, compared to the cost of 70 epochs per minute for 100 points reported in Sec.~\ref{sec_case2_RandD}, represents an increase in calculation speed of about 16\%. However, as already pointed out, the batch size of 10 did not provide the correct solution. 
\begin{figure}
	\centering
	\includegraphics[width=8cm]{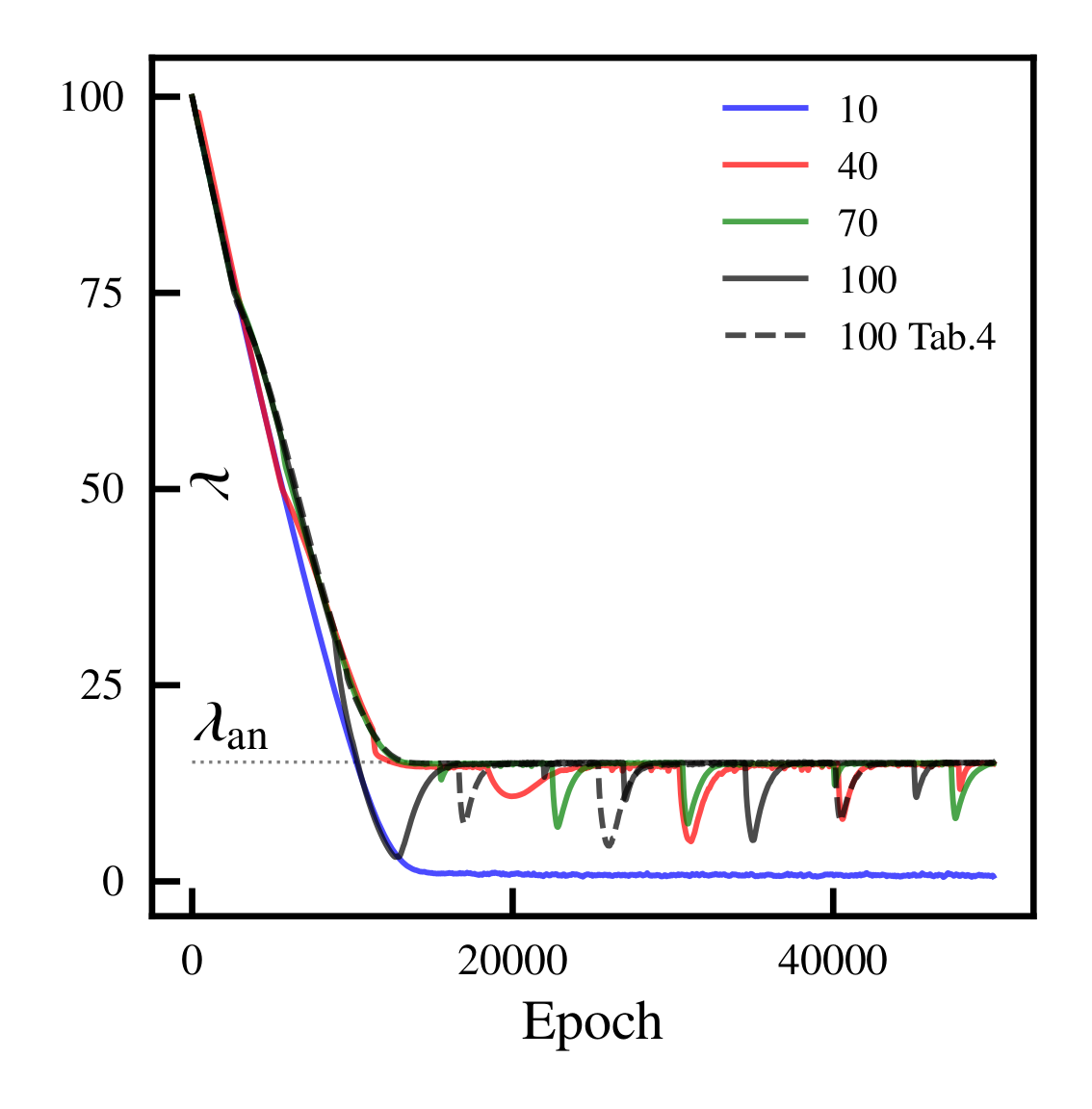}
	\caption{Convergence of $\lambda$ for batches of $N_\mathrm{c} \in {10, 40, 70, 100}$ randomly selected collocation points for the nonlocal cantilever beam with a learning rate of $10^{-2}$. The results for $N_\mathrm{c}=100$ are taken from Tab.~\ref{tab_Nonlocal_CF} and Fig.~\ref{fig_CF_HR}.}
	\label{Fig_batches}
\end{figure}

\begin{table}
	\begin{center}
		\begin{tabular}{|l|r|r|r|r|r|}
			\hline
			$N_\mathrm{c}$		  & 10		& 40	&70		&100 	&100 (Tab.\ref{tab_Nonlocal_CF})    \\
			\hline
			Predicted $\lambda$   & 0.7261   & 15.0413 & 15.1144 & 15.1411 & 15.1182 \\
			Difference (\%) 	  & 1992.728 &  1.0238 &  0.5353 &  0.3580 & 0.4741   \\
			Best epoch			  & \numprint{34371} & \numprint{49666} & \numprint{46024} & \numprint{44732} & \numprint{49774}  \\
			\hline
			min. $L_\mathrm{tot}$ & 246.6275 & 1.0948 & 0.2408 & 0.2599 & 0.5144 \\
			\hline
			
		\end{tabular}
		\caption{Number of randomly selected collocation points ($N_\mathrm{c}$), predicted eigenvalue ($\lambda$), difference from the analytical eigenvalue ($\lambda_\mathrm{an}=15.1953$), and minimal total loss obtained for a high learning rate in the nonlocal cantilever beam. Number of training epochs: \numprint{50000}, $\lambda_\mathrm{ini}=100$.}
		\label{tab_Nonlocal_CF_batch}
	\end{center}
\end{table}

\section{Influence of Initial Points}
\label{sec:app6_initial}
This section examines the effect of different initial values of $\lambda$ to demonstrate the convergence properties of the proposed method. A nonlocal cantilever beam is considered. To provide better insight, two learning rates were tested: $10^{-2}$ and $10^{-3}$. The following initial values were selected: $\lambda_\mathrm{ini} \in \left\lbrace20, 30, 60, 75, 80, 100\right\rbrace $. Except for one case, all simulations started from the same set of random weights. The obtained results are shown in Fig.~\ref{Fig_lambda_all_ini}. Although requiring more epochs to converge, the learning rate of $10^{-3}$ resulted in a more stable asymptotic value, evident from the absence of jumps observed for the learning rate of $10^{-2}$. Nevertheless, for one starting point ($\lambda_\mathrm{ini}=60$), convergence to a trivial mode was observed. Allowing for different random weight initialization (case 60r, dashed–dot line) enabled correct convergence. Convergence to a trivial mode was also observed for $\lambda_\mathrm{ini}=75$ and learning rate $10^{-2}$. In the case of $\lambda_\mathrm{ini}=20$ and learning rate $10^{-2}$, temporary convergence to the trivial mode occurred, followed by correct convergence afterward. This behavior can be attributed to the higher learning rate, for which it is typically more difficult to become trapped in a local minimum.

To conclude, the method exhibits the typical characteristics of stochastic optimization, where convergence to the global minimum is not guaranteed and is usually ensured by repeating the training with different random weight initializations. In this particular case, 8 out of 10 simulations converged to the correct optimum.

\begin{figure}
	\centering
	\includegraphics[width=8cm]{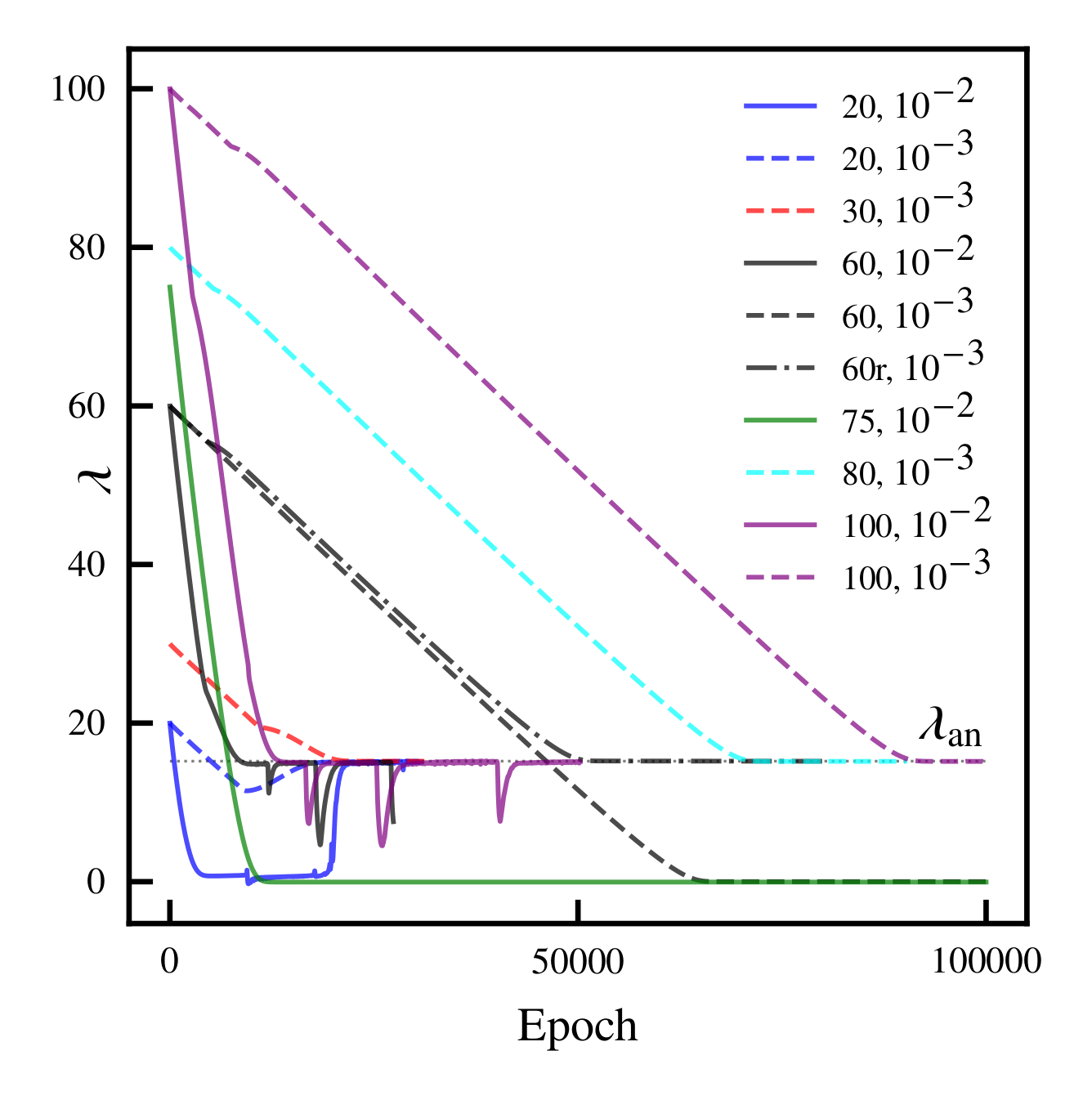}
	\caption{Convergence of $\lambda$ for different initial points $\lambda_\mathrm{ini} \in \left\lbrace20, 30, 60, 75, 80, 100\right\rbrace$ and learning rates ($10^{-2}$/solid lines and $10^{-3}$/dashed lines) in the nonlocal cantilever beam. All cases start with the same random weights, except for case 60r (dashed-dot line), which uses random weight initialization.}
	\label{Fig_lambda_all_ini}
\end{figure}

\newpage
%\bibliographystyle{elsarticle-num}
%\bibliography{bibliography.bib}

\end{document}